\documentstyle[epsf,11pt]{article}

\newtheorem{theorem}{Theorem}

\newtheorem{lemma}{Lemma}

\newtheorem{corollary}[theorem]{Corollary}
\newenvironment{proof}%
 {\par\noindent{\underline{Proof} \quad}}{\hfill$\Box$\bigskip}
 {\par\noindent{\underline{Proof} of the theorem\quad}}{\hfill$\Box$\bigskip}
 {\par\noindent{\underline{Proof} of the lemma\quad}}{\hfill$\Box$\bigskip}

\newenvironment{remark}%
 {\par\smallskip\noindent{\underline{{\it Remark}} \quad}}{\par\smallskip}
 {\par\smallskip\noindent{\underline{{\it Fact}} \quad}}{\par\smallskip}
\newenvironment{example}%
 {\par\smallskip\noindent{\underline{{\it Example}} \quad}}{\par\smallskip}
 {\par\smallskip\noindent{{\it Assumotion} \quad}}{\par\smallskip}
 {\par\smallskip\noindent{{\it Condition} \quad}}{\par\smallskip}

\newcommand{\ol}{\overline}

\newcommand{\tr}{{\rm tr}}
\newcommand{\Tr}{{\rm Tr}}
\newcommand{\R}{{\bf R}}
\newcommand{\C}{{\bf C}}

\newcommand{\rank}{{\rm rank}}
\newcommand{\abs}{{\rm abs}}

\newcommand{\CR}{{\rm CR}}

\newcommand{\rgl}{\rangle}
\newcommand{\lgl}{\langle}

\begin{document}
\begin{center}

\vspace*{2mm}
{\Large
A new approach  to the Cramer-Rao type bound \\
of  the pure state model }\\
\vspace{10mm}
Keiji Matsumoto
\footnote{
Department of Mathematical Engineering and Information Physics \\
University of Tokyo, Bunkyo-ku,Tokyo 113, Japan }
\begin{abstract}
In this paper,   new methodology -- direct approach -- 
 for the determination of the attainable CR type bound
of the pure state model, is proposed and successfully applied to 
the wide variety of pure state models,
for example,
 the 2-dimensional arbitrary model, 
the coherent model with arbitrary dimension.
When the weight matrix is $SLD$ Fisher information,
the bound is determined for arbitrary pure state models. 
Manifestation  of complex structure in the Cramer-Rao type 
bound is also discussed.
\end{abstract}

        
\end{center}

\section{Introduction}
\label{sec:introduction}
The  quantum estimation theory deals with 
determination of  the density operator of the given physical system
from the data obtained in the experiment.
For simplicity, 
it is assumed that 
a state  belongs to a certain subset 
${\cal M}
=\{\rho(\theta)|\theta\in \Theta\subset {\R}^m\}$ 
of the space of the states,
which is called {\it model},
and that
the true value of the finite dimensional parameter $\theta$ 
is left to be estimated statistically.
In this paper, we restrict ourselves to pure state model case, where
 ${\cal M}$ is  a subset of 
the space ${\cal P}_1$ of pure states in
$d$-dimensional Hilbert space ${\cal H}$ $(d\leq\infty)$. 
For example, ${\cal M}$ is 
a set of spin states with given wave function part and unknown spin part.

In the classical estimation
(throughout the paper, `classical estimation' means the estimation theory of
probability distribution),
the mean square error is often used as a measure of error of the estimate,
and 
the Cramer-Rao inequality assures that
the inverse of so-called Fisher information matrix
is the tight lower bound of
covariance matrices of locally unbiased estimator
(Ref.$\cite{Lehmann:1983}$).

Analogically, in the quantum estimation theory,
in 1967, Helstrom showed that in the faithful state model,
the covariance matrix is larger than or equal to the inverse of 
SLD Fisher information matrix, and that in the $1$-dimensional faithful model,
the bound is attainable \cite{Helstrom:1967}\cite{Helstrom:1976}.

On the other hand, in the multi-dimensional model,
it is proved that there is no matrix 
which makes attainable lower bound of covariance matrix,
because of non-commutative nature of quantum theory.
Hence, the measure of the error of the estimate which is often used 
is $\Tr GV_{\theta}[M]$, where
$V_{\theta}[M]$ denotes  
the covariance matrix of the locally unbiased measurement  $M$ at $\theta$
and $G$ is a {\it weight matrix}, 
or an arbitrary given $m\times m$ positive symmetric real matrix.
The infimum of  is
said to be {\it attainable} or  {\it achievable}
Cramer-Rao (CR) type bound of the model 
at $\theta$ with weight matrix $G$,
and to determine the attainable CR type bound
long had been one of the main topics in this field, 
and is solved only for
the several specific models,
because $\Tr GV_{\theta}[M]$ is a functional of probability valued
measure, or pair of infinite number of operators in the infinite
dimensional Hilbert space.

Yuen, Lax and Holevo
found out the attainable CR type bound 
of the Gaussian state model,
which is a faithful 2-dimensional model obtained by superposition of
coherent states by Gaussian kernel\cite{YuenLax:1973}\cite{Holevo:1982}.
Nagaoka and Hayashi  calculated the attainable CR type bound of
the faithful faithful spin-$1/2$ model\cite{Nagaoka:1989}\cite{Hayashi:1997}.
Fujiwara and Nagaoka determined the bound 
for the 1-dimensional pure state model
and the 2-dimensional coherent model,
which is the pure-state-limit of the Gaussian model.
\cite{FujiwaraNagaoka:1995}\cite{FujiwaraNagaoka:1996}.

All of their works are based on a methodology,
which we call {\it indirect approach} hereafter;
First one somehow  find an auxiliary bound which is not generally attainable
and then proves it to be attained in the specific cases.

In the approach in this paper,  called {\it direct approach} 
in contrast with {\it indirect approach},
we reduce the problem to the minimization of the functional
of the finite numbers of the finite dimensional vectors.

The methodology is successfully applied to 
the general 2-dimensional pure state model,
and coherent model with arbitrary dimension.
These are relatively general category in comparison with the cases
treated by other authors. 
Also, when the weight matrix is SLD Fisher information matrix,
which will be defined in somewhere in the paper,
the bound is calculated for arbitrary pure state models.

As a by-product, we have  rather paradoxical corollary,
which asserts that even for `non-commutative cases',
simple measurement attains the lower bound.

The paper is organized as follows.
In section 2 and 3,
 basic concepts of the quantum estimation theory are introduced.
In section 4, the commuting theorem, which plays key role in the foundation of 
the direct approach, is presented and is applyed to the characterization
the quasi-classical model, in which non-commutative nature
of the theory is not apparent.
We formulate the problem in the non-quasi-classical models in section 5.
Our new methodology, direct approach, 
is introduced in section 6 and 7,
and is applied to the 2-dimensional pure state model
and the coherent model in section 8 and 11 respectively.
In section 9, we consider informational correlation between the parameters,
 and the attainable CR type bound for the direct sum of the models.
The manifestation of the quantum structure,
together with the minimization of
the minimum of $\Tr J^S(\theta)V_{\theta}[M]$, is discussed in section 10.

\section{Locally unbiased measurement}
\label{sec:unbiased}
Let $\sigma({\R}^m)$ be a $\sigma$- field in the space ${\R}^m$.
Whatever measuring apparatus is used to produce 
the estimate $\hat\theta$ of the true value of the parameter $\theta$,
the probability that the estimate $\hat{\theta}$
lie in a particular measurable set $B$
in ${\bf R}^m$  will be given by
\begin{eqnarray}
{\rm Pr}\{ \hat{\theta} \in B|\theta \} =\tr \rho( \theta )M(B)
\label{eqn:pdm}
\end{eqnarray}
when $\theta$ represents the true value of parameter.
Here $M$ is a mapping
of a measurable set $B\in \sigma({\R}^m) $ 
to non-negative Hermitian operators on ${\cal H}$, such that
\begin{eqnarray}
&&M(\phi)=O,M({\bf R}^m)=I,\nonumber\\
&&M(\bigcup_{i=1}^{\infty} B_i)
=\sum_{i=1}^{\infty}M(B_i)\;\;(B_i\cap B_j=\phi,i\neq j),
\label{eqn:pom}
\end{eqnarray}
(see Ref.{\rm \cite{Helstrom:1976}},p.53 
and Ref.{\rm \cite{Holevo:1982}},p.50.).
 $M$ is called a {\it generalized measurement} or 
{\it measurement}, 
because
there is a corresponding measuring apparatus
to any $M$ satisfying $(\ref{eqn:pom})$
\cite{Ozawa:1984}\cite{steinspring:1955}.
A measurement $E$ is said to be {\it simple} if $E$ is projection valued.

A generalized measurement $M$ is called an {\it unbiased measurement} 
in the model ${\cal M}$,
if $E_\theta[M]=\theta$ holds 
for all $\theta\in\Theta$, i.e.,
\begin{equation}
 \int\hat\theta^j \tr\rho(\theta)\, M((d\hat{\theta})
=\theta^j,\quad(j=1,\cdots,m).
\label{eqn:unbiasedness}
\end{equation}
Differentiation yields
\begin{equation}
 \int\hat\theta^j \tr\frac{\partial\rho(\theta)}{\partial \theta^k}\, 
  M(d\hat{\theta})
   =\delta^j_k,\quad(j,k=1,\cdots,m).
\label{eqn:local_unbiasedness}
\end{equation}
If (\ref{eqn:unbiasedness}) and (\ref{eqn:local_unbiasedness}) hold at a 
some $\theta$, $M$ is said to be {\it locally unbiased} at $\theta$.
Obviously, $M$ is unbiased iff $M$ is locally unbiased 
at every 
$\theta\in\Theta$.

As a measure of error of a locally unbiased measurement $M$,
we employ
the covariance matrix with respect to $M$ at the state $\rho_\theta$,
$V_{\theta}[M]=[v^{jk}_{\theta}]\in{\bf R}^{m\times m}$, where 
\begin{equation}
 v^{jk}_{\theta}=\int(\hat\theta^j-\theta^j)(\hat\theta^k-\theta^k)
                 \tr \rho(\theta)M(d\hat{\theta}). 
\label{eqn:covariant}
\end{equation}
We often abbreviated notation $V[M]$ for $V_{\theta}[M]$ 
when it is not confusing.
The problem treated in this note is to find a lower bound for
$V_{\theta}[M]$.

Only locally unbiased measurements are treated from now on,
because of the following reason.
Given $N$ copies of the system,
we apply a proper measurement to the the first $pN$ copies,
and the true value of parameter is known to lie in certain
$\epsilon$-ball centered at $\theta_0$ 
with the probability $\sim 1- e^{- a/\epsilon^2 N}$.
Therefore, applying the `best' locally unbiased measurement 
at $\theta_0$ to the $(1-p)N$ copies, 
we can achieve the efficiency arbitrarily close to
that of the `best' locally unbiased measurement at $\theta$,
in the sense of the first order asymptotics.

\section{CR bound by SLD Fisher information matrix}
\label{sec:sld}
In 1995,  Fujiwara and Nagaoka \cite{FujiwaraNagaoka:1995}
defined SLD Fisher information for pure state models.
Here,
we try another definition which is adequate for our direct approach.

Analogically to the classical estimation theory, 
in the quantum estimation theory,
 we have the following {\it SLD CR inequality},
which is proved for the exact state model 
by Helstrom \cite{Helstrom:1967}\cite{Helstrom:1976},
and is proved for the pure state model by Fujiwara and Nagaoka
\cite{FujiwaraNagaoka:1995}:
\begin{eqnarray}
V_{\theta}[M]\geq(J^S(\theta))^{-1},
\label{eqn:mpCR}
\end{eqnarray}
{\it i.e.}, $V_{\theta}[M]-(J^S(\theta))^{-1}$ is non-negative definite.
Here $J^S(\theta)$, called {\it SLD Fisher information matrix}, 
is defined by
\begin{eqnarray}
J^S(\theta)\equiv [{\rm Re}\lgl l_i(\theta)|l_j(\theta)\rgl],
\nonumber
\end{eqnarray}
where  the notations 
$|\l_i(\theta)\rgl\:(i=1,...,m)$ are defined afterward.

The inequality $(\ref{eqn:mpCR})$ is of special interest,
because $J^{S-1}(\theta)$  is the one of the best bounds 
in the sense of the following theorem, 
which will be proved in the section $\ref{sec:nonclassical}$.
\begin{theorem}
Letting $A$ be a real hermitian matrix which is larger than $J^{S-1}$,
that is, $A>J^{S-1}$, there exists such an unbiased estimator $M$
that $V[M]$ is not smaller than $A$.
\label{theorem:sldbest} 
\end{theorem}
To define the notations 
$|\l_i(\theta)\rgl\:(i=1,...,m)$ and to prove  the SLD CR inequality,
 we introduce some basic notations.
$\tilde{\cal H}$ is a set of vectors with unit length,
\begin{eqnarray}
\tilde{\cal H}=\{|\phi\rgl\:|\:|\phi\rgl\in{\cal H},\lgl\phi|\phi\rgl=1\}.
\nonumber
\end{eqnarray}
${\cal P}_1$ denotes 
the totality of density operators of pure states in ${\cal H}$.
A map $\pi$ from $\tilde{\cal H}$ to ${\cal P}_1$ is defined by
\begin{eqnarray}
\pi(|\phi\rgl) \equiv |\phi\rgl\lgl\phi|.
\nonumber
\end{eqnarray}
For the manifold 
${\cal N}=\{ |\phi(\theta)\rgl\, |\, \theta\in\Theta\subset \R^m\}$
in $\tilde{\cal H}$, $\pi({\cal N})$ is defined to be a manifold 
in ${\cal P}_1$ such that
\begin{eqnarray}
\pi({\cal N})
=\{\rho(\theta)\, |\,
  \rho(\theta)= \pi(|\phi(\theta)\rgl),\,|\phi(\theta)\rgl\in{\cal N}\}.
\nonumber
\end{eqnarray}
Through out the paper, we only treat with the pure state model ${\cal M}$ 
which writes ${\cal M}=\pi({\cal N})$ 
for a manifold ${\cal N}$ in  ${\cal P}_1$.

The {\it horizontal lift} 
$|l_X\rgl$ of a tangent vector
$X\in {\cal T}_{\rho(\theta)}({\cal M})$ to $|\phi(\theta)\rgl$,
is an element of ${\cal H}$ which satisfies
\begin{eqnarray}
X\rho(\theta)=
\frac{1}{2}(|l_X\rgl\lgl\phi(\theta)|+|\phi(\theta)\rgl\lgl l_X|),
\label{eqn:sld:lift}
\end{eqnarray}
and
\begin{eqnarray}
\lgl l_X|\phi(\theta)\rgl=0.
\label{eqn:sld:horizontal}
\end{eqnarray}
Here, 
$X$ in the left hand side $(\ref{eqn:sld:lift})$ of is to be understood as
a differential operator.
We use the symbol $|l_i(\theta)\rgl$ to denote a horizontal lift of 
$\partial_i\in {\cal T}_{\rho(\theta)}({\cal M})$. 

Notice that 
$span_{\bf R}\{|l_i\rgl\;|\: i=1,...,m\}$ is a representation of 
${\cal T}_{\rho(\theta)}({\cal M})$ 
because of unique existence of the horizontal lift to $|\phi(\theta)\rgl$
which is proved as follows.
Application of a differential operator $X$ 
to the both sides of $\rho(\theta)=\rho^2(\theta)$
yields 
\begin{eqnarray}
X\rho(\theta)=
(X\rho(\theta))|\phi(\theta)\rgl\lgl\phi(\theta)|
+|\phi(\theta)\rgl\lgl\phi(\theta)|(X\rho(\theta))
\label{eqn:sld:rho^2}
\end{eqnarray}
and therefore $|l_X\rgl$ is given by 
$(\frac{1}{2}X\rho(\theta))|\phi(\theta)\rgl$.
Actually, taking trace of both sides of $(\ref{eqn:sld:rho^2})$,
it is shown that $(X\rho(\theta))|\phi(\theta)\rgl$ satisfies 
$(\ref{eqn:sld:horizontal})$. 
To prove the uniqueness,
it suffices to show that $|l\rgl=0$ if $\lgl l|\phi(\theta)\rgl=0$ and 
\begin{eqnarray}
0=|l\rgl\lgl\phi(\theta)|+|\phi(\theta)\rgl\lgl l|
\label{eqn:sld:lplp=0}
\end{eqnarray}
holds true.
Multiplication of $|\phi(\theta)\rgl$ to the both sides of 
$(\ref{eqn:sld:lplp=0})$ proves the statement.

Fujiwara and Nagaoka defined SLD Fisher information matrix $J^S(\theta)$
by using
the symmetric logarithmic derivative (SLD)
of the parameter $\theta^i$ is a hermitian matrix $L^S_i(\theta)$ 
which satisfies
\begin{eqnarray}
\partial_i \rho(\theta)
=\frac{1}{2}(L_i^S(\theta)\rho(\theta)+\rho(\theta)L^S_i(\theta)).
\label{eqn:defsld}
\end{eqnarray}
Using SLD, the horizontal lift of $\partial_i$ to 
${\cal T}_{|\phi(\theta)\rgl}(\tilde{\cal H})$ writes
$|l^i(\theta)\rgl=L^S_i(\theta)|\phi(\theta)\rgl$.
$J^S(\theta)$ is called  SLD Fisher information matrix because 
$J^S(\theta)$ writes 
\begin{eqnarray}
J^S(\theta)=[{\rm Re}\:\tr\rho(\theta) L_i^S(\theta) L_j^S(\theta)].
\end{eqnarray}
SLD defined by $(\ref{eqn:defsld})$ has the arbitrariness
which corresponds to the kernel of $\rho(\theta)$, and
Fujiwara and Nagaoka \cite{FujiwaraNagaoka:1995}
 showed that $J^S(\theta)$ is uniquely defined
regardless this arbitrariness.
Notice that 
in our framework, uniqueness of SLD Fisher information matrix is trivial.

We define {\it estimation vector} $|x^i[M,|\phi(\theta)\rgl]\rgl$ 
of the parameter $\theta^i$ 
by a measurement $M$ at $|\phi(\theta)\rgl$, by
\begin{eqnarray}
|x^i[M,|\phi(\theta)\rgl]\rgl\equiv 
\int (\hat{\theta}^i -\theta^i)M(d\hat{\theta})|\phi(\theta)\rgl.
\nonumber
\end{eqnarray}
An estimation vector $|x^i[M,|\phi(\theta)\rgl]\rgl$ 
is said to be {\it locally unbiased}
iff $M$ is locally unbiased.
The local unbiasedness conditions for estimating vectors writes
\begin{eqnarray}
\lgl x^i[M,|\phi(\theta)\rgl]|\phi(\theta)\rgl&=&0,\\
{\rm Re}\lgl x^i[M,|\phi(\theta)\rgl]| l_j(\theta)\rgl&=&\delta^i_j
\:(i,j=1,...,m).
\label{eqn:sld:hunbiased}
\end{eqnarray}
Often, we omit the argument  $\theta$ in
$| l_j(\theta)\rgl,|\phi(\theta)\rgl, \rho(\theta)$, and $J^S(\theta)$
and denote them simply by $| l_j\rgl,|\phi\rgl, \rho, J^S$.
Also, $|x^i[M,|\phi(\theta)\rgl]\rgl$ is denoted simply 
by $|x^i\rgl$ with the arguments $M,|\phi(\theta)\rgl$ left out,
so far as no confusion is caused. 

We denote the ordered pair of vectors 
\begin{eqnarray}
[|x^1\rgl,|x^2\rgl,...|x^m\rgl]
\nonumber
\end{eqnarray}
and 
\begin{eqnarray}
[|l'_1\rgl,|l'_2\rgl,...|l'_m\rgl]
\nonumber
\end{eqnarray}
by ${\sf X}$ and ${\sf L}$ respectively.
Then, the unbiasedness conditions  $(\ref{eqn:sld:hunbiased})$ 
writes
\begin{eqnarray}
{\rm Re}{\sf X}^*{\sf L}\equiv {\rm Re}[\lgl x^i|l^j \rgl]=I_m,
\label{eqn:lagrange:restriction1}
\end{eqnarray}
where $I_m$ is the $m\times m$ unit matrix.
The SLD Fisher information matrix $J^S$ writes
\begin{eqnarray}
J^S={\rm Re}{\sf L}^*{\sf L}.
\nonumber
\end{eqnarray}
The imaginary part of ${\sf L}^*{\sf L}$
is denoted by $\tilde{J}$.

Now, we are in the position to  prove 
SLD CR inequality.
\begin{lemma}
Following two inequalities are valid:
\begin{equation}
  V[M]\ge {\rm Re}{\sf X}^*{\sf X}.
\label{eqn:inVZ}
\end{equation}
\begin{equation}
  V[M]\ge {\sf X}^*{\sf X}.
\label{eqn:inVZ2}
\end{equation}
\label{lemma:VZ}
\end{lemma}
\begin{lemma}
\begin{eqnarray}
{\rm Re}{\sf X}^*{\sf X}\ge J^{S-1}
\label{eqn:ZJ}
\end{eqnarray}
 holds. 
The equality is valid iff
\begin{eqnarray}
|x^j\rgl =\sum_k [J^{S-1}]^{j,k}|l_k\rgl,
\nonumber
\end{eqnarray}
or, equivalently,
\begin{eqnarray}
{\sf X}={\sf L}J^{S-1}\equiv 
\left[\sum_k [J^{S-1}]^{j,k}|l_k\rgl,\:j=1,...,m\right]
\label{eqn:xx=J}
\end{eqnarray}
\label{lemma:ZJ}
\end{lemma}
They are proved in almost the same manner as the strictly positive case
(see Ref.\cite{Holevo:1982} p.88 and p.274 respectively).
Lemmas $\ref{lemma:VZ}$-$\ref{lemma:ZJ}$ lead to 
the SLD CR inequality $(\ref{eqn:mpCR})$.
\begin{theorem}(Fujiwara and Nagaoka\cite{FujiwaraNagaoka:1995})
SLD Fisher information gives a lower bound of covariance matrix
of an unbiased measurement, {\it i.e.},
$(\ref{eqn:mpCR})$ holds true.
\end{theorem}

The SLD CR inequality $(\ref{eqn:mpCR})$ looks quite analogical to
CR inequality in classical estimation theory.
However, as is found out in the next section,
the equality does not generally establish.

\section{The commuting theorem and the quasi-classical model}
\label{sec:classical}
In this section,
the necessary and sufficient condition 
for the equality in the SLD CR inequality to establish is studied. 
Fujiwara has proved the following theorem \cite{Fujiwara}.
\begin{theorem}(Fujiwara \cite{Fujiwara})
The equality in the SLD CR inequality establishes iff
 SLDs $\{L^S_i|i=1,...,m\}$ can be chosen so that
\begin{eqnarray}
[L^S_i, L^S_j]=0,\,\, (i,j=1...,m).
\nonumber
\end{eqnarray}
\label{theorem:fjcl}
\end{theorem}
We prove another necessary and sufficient condition which is much easier
to check for given models, by use of
the following {\it commuting theorem}, which
plays key role in our direct approach.
\begin{theorem}
If there exists a unbiased measurement $M$ such that
\begin{eqnarray}
&&|x^i\rgl=\int (\hat\theta^i-\theta^i)M(d\hat\theta)|\phi\rgl,\nonumber\\
&&V[M]={\rm Re}{\sf X}^*{\sf X},
\label{eqn:VZ}
\end{eqnarray}
then,
\begin{eqnarray}
{\rm Im}{\sf X}^*{\sf X}=0
\label{eqn:imxx=0}
\end{eqnarray}
holds true.
On the other hand, if $(\ref{eqn:VZ})$ holds true, then
there exists  such a simple, or projection valued, 
unbiased measurement $E$ that
$(\ref{eqn:VZ})$ holds and
\begin{eqnarray}
&&E(\{\hat\theta_{\kappa}\})E(\{\hat\theta_{\kappa}\})
    = E(\{\hat\theta_{\kappa}\}),\nonumber\\
&&E(\{\hat\theta_{0}\}) = E_0,\nonumber\\
&&E\left( {\R}^m/\bigcup_{\kappa=0}^m \{\hat\theta_{\kappa}\} 
                                                 \right)=0,
\label{eqn:commes}
\end{eqnarray}
for some 
$\{\hat\theta_{\kappa}|\hat\theta_{\kappa}\in {\R}^m,\kappa = 0,...,m+1\}$,
where $E_0$ is a projection onto orthogonal complement subspace
of $span_{\C}\{{\sf X}\}$.
\label{theorem:commute}
\end{theorem}
\begin{proof}
If $(\ref{eqn:VZ})$ holds,
inequality $(\ref{eqn:inVZ2})$ in lemma $\ref{lemma:VZ}$
leads to
\begin{eqnarray}
{\rm Re}{\sf X}^* {\sf X}\ge {\sf X}^*{\sf X},\nonumber
\end{eqnarray}
or
\begin{eqnarray}
0 \ge  i{\rm Im}{\sf X}^*{\sf X},\nonumber
\end{eqnarray}
which implies  ${\rm Im}{\sf X}^*{\sf X}=0$.

Conversely, Let us assume that  $(\ref{eqn:imxx=0})$ holds true.
Applying Schmidt's orthogonalization to
$\{|\phi\rgl, |x^1\rgl,..., |x^m\rgl\}$ and normalizing the product of
orthogonalization,
we obtain the orthonormal system $\{|b^i\rgl\;|\; i=1,...,m+1\}$
of vectors such that,
\begin{eqnarray}
|x^i\rgl=\sum_{j=1}^{m+1} \lambda^i_j|b^j\rgl,
\:\:\exists {\lambda^j_i}\in{\bf R}, i=1,...,m,j=1,...,m+1.
\nonumber
\end{eqnarray}
Letting  $O=[o^i_j]$ be a $(m+1)\times (m+1)$ real orthogonal matrix
such that
\begin{eqnarray}
\lgl\phi|\sum_{j=1}^{m+1} o^i_j |b^j\rgl\neq 0,
\nonumber
\end{eqnarray}
and denoting $\sum_{j=1}^{m+1} o^i_j |b^j\rgl$ by $|b'^{i}\rgl$,
we have
\begin{eqnarray}
|x^i\rgl
&=&\sum_{j=1}^{m+1} \lambda^i_j\sum_{k=1}^{m+1} o^k_j |b'^k\rgl\nonumber\\
&=&\sum_{k=1}^{m+1}
        \left(\sum_{j=1}^{m+1} \lambda^i_j o^k_j \right)|b'^k\rgl\nonumber\\
&=&\sum_{k=1}^{m+1}
        \frac{\sum_{j=1}^{m+1} \lambda^i_j o^k_j }{\lgl b'^k|\phi\rgl}
                |b'^k\rgl\lgl b'^k|\phi\rgl.
\nonumber
\end{eqnarray}
Therefore, noticing that the system $\{|b'^i\rgl\:|\: i=1,...,m+1\}$ of vectors
is orthonormal,
we obtain an unbiased measurement which satisfies $(\ref{eqn:commes})$
as follows:
\begin{eqnarray}
&&\hat\theta_{\kappa}=
\frac{\sum_{j=1}^{m+1} \lambda^i_j o^{\kappa}_j }{\lgl b'^{\kappa}|\phi\rgl},
\:\:\kappa =1,...,m+1,\nonumber\\
&&\hat\theta_{0}=0,\nonumber\\
&&E(\{\hat\theta_{\kappa}\})=|b'^{\kappa}\rgl\lgl b'^{\kappa}|,\:\:
\kappa=1,...,m+1,\nonumber\\
&&E(\hat\theta_{0})=
I_{\cal H}-\sum_{\kappa=1}^{m+1}|b'^{\kappa}\rgl\lgl b'^{\kappa}|,\nonumber\\
&&E(\R^m/\{\hat\theta_{0},...,\hat\theta_{m+1} \})=0.
\nonumber
\end{eqnarray}
Here, $I_{\cal H}$ is the identity in ${\cal H}$.
\end{proof}
\begin{theorem}
The equality in the SLD CR inequality establishes iff
\begin{eqnarray}
{\rm Im}{\sf L}^*{\sf L}=0
\label{eqn:lsls0}
\end{eqnarray}
$\lgl l_j|l_i\rgl$ is real for any $i,j$.
When the equality establishes, that bound is achieved 
by a simple measurement, i.e., a projection valued measurement.
\label{theorem:purelsls0}
\end{theorem}
\begin{proof}
If the equality establishes,
by virtue of lemma $\ref{lemma:VZ}$-$\ref{lemma:ZJ}$,
we have $(\ref{eqn:imxx=0})$ and 
$(\ref{eqn:xx=J})$, which lead directly to $(\ref{eqn:lsls0})$.

Conversely, if ${\rm Im}\lgl l_j|l_k\rgl =0$ for any $j,k$,
by virtue of commuting theorem, 
there exists such a simple measurement $E$  that
\begin{eqnarray}
\sum_k [J^{S-1}]^{j,k}|l_k\rgl=\int (\hat\theta-\theta)E(d\hat\theta)|\phi\rgl.
\nonumber
\end{eqnarray}
Elementary calculations show that the covariance matrix of this measurement
equals  $J^{S-1}$.
\end{proof}
Our theorem is equivalent to Fujiwara's one,
because 
by virtue of commuting theorem,
$\lgl l_j|l_i\rgl$ is real iff there exist such SLDs that
$L_i^S$ and $L_j^S$ commute for any $i,j$.
However, our condition is much easier to be checked,
because to check Fujiwara's condition, you must calculate 
all the possible SLDs, for the SLD is not unique.
In addition,  SLD  is  much harder to calculate
than horizontal lift. 
When the model has only one -dimensional, we have the following corollary
of theorem $\ref{theorem:purelsls0}$

\begin{corollary}
when a manifold ${\cal M}$ 
is one-dimensional, the inverse of SLD Fisher information matrix
is always attainable
by a simple measurement.
\label{theorem:1pCR}
\end{corollary}
\begin{remark}
Often, a model is defined by an initial state and  generators,
\begin{eqnarray}
{\cal N}&\equiv&
    \{|\phi(\theta)\rgl\,|\,
          |\phi(\theta_0)\rgl=|\phi_0\rgl,\,
          \partial_i|\phi(\theta)\rgl=iH_i(\theta)|\phi(\theta)\rgl,\,
          \theta\in\Theta\subset\R^m\},\nonumber\\
{\cal M}&\equiv&\pi({\cal N}).\nonumber
\end{eqnarray}
Then, $\lgl l_j|l_i\rgl$ is real iff
$\lgl\phi(\theta)|[H_i(\theta),H_j(\theta)]|\phi(\theta)\rgl=0$,
which is equivalent to the existence of  
generators which commute with each other, $[H_i(\theta),H_j(\theta)]=0$
by virtue of the commuting theorem.
\end{remark}
Putting the remark and the  Fujiwara's theorem  together, 
we may metaphorically say 
that the equality in the inverse of SLD Fisher information matrix
 is attainable iff any two parameters 
`commute' at $\theta$.
Throughout the paper, we say that a manifold ${\cal M}$ is
{\it quasi-classical} at $\theta$  iff $\lgl l_j|l_i\rgl$ is real at $\theta$.
The following remark describes another `classical' aspect of 
the condition ${\rm Im}\lgl l_j|l_i\rgl=0$.
\begin{example}
When the model ${\cal M}$ is given by 
\begin{eqnarray}
{\cal M}=\{\rho(\theta)\:|\: \rho(\theta)=\pi(|\phi(\theta)\rgl),\:
                |\phi(\theta)\rgl \mbox{is an element of real Hilbert space}\},
\nonumber
\end{eqnarray}
the model is quasi-classical at any point in ${\cal M}$.
\end{example}
As is illustrated in this example, 
when th model ${\cal M}$ is quasi-classical at $\theta_0$,
a state vector $|\phi(\theta)\rgl$
behaves like an element of real Hilbert space around  $\theta_0$,
and the state vector's phase parts  don't change around $\theta_0$
at all.

\section{Non-quasi-classical cases}
\label{sec:nonclassical}
As was concluded,
the equality in the SLD CR inequality establishes 
only when the model is quasi-classical, 
and there is not
any better bound than the inverse of SLD Fisher information matrix,
as in  theorem $\ref{theorem:sldbest}$, which is
straightforwardly derived from the following lemma,
which is proved in the appendix A.
\begin{lemma}
For any $i$,
\begin{eqnarray}
\inf \left\{\left. [V_{\theta}[M]]_{ii}\, \right|\, 
          \mbox{ $M$ is locally unbiased at $\theta$}\right\}
=\left[J^{S-1}(\theta)\right]^{ii}
\nonumber
\end{eqnarray}
\label{lemma:eivei}
\end{lemma}
In general case, therefore, we must give up 
to find a matrix which makes attainable lower bound of $V[M]$,
and instead, we try to determine 
\begin{eqnarray} 
\CR(\theta, G, {\cal M})\equiv\inf\{\Tr GV\: |\: V\in{\cal V}_{\theta}({\cal M})\}
\label{eqn:crtype}
\end{eqnarray}
for an arbitrary nonnegative symmetric real matrix $G$,
where  ${\cal V}_{\theta}({\cal M})$ (, or in short, ${\cal V}$,) 
is the region of the map $V_{\theta}[*]$ from unbiased estimators 
to $m \times m$ real positive symmetric matrices.
$\CR(\theta, G, {\cal M})$ is the {\it attainable CR type bound},
and we often use abbreviated notations such as
$\CR(\theta, G)$, $\CR(G)$ .

To make the estimational meaning of $(\ref{eqn:crtype})$ clear, let us
restrict ourselves to the case when  $G$ is $diag(g_1,g_2...,g_m)$. 
Then, the attainable CR type bound is
nothing but 
 the weighed sum of the covariance of the estimation of $\theta^i$. 
 If one needs to know, for example, $\theta^1$ 
more precisely than other parameters,
then he set  $g_1$ larger than any other $g_i$,
and choose a measurement which achieves the attainable CR type bound.

Notice that 
\begin{eqnarray}
&&\inf \left\{\left. \sum_{i=1}^m g_i [V_{\theta}[M]]_{ii}\, \right|\, 
          \mbox{ $M$ is locally unbiased at $\theta$}\right\}\nonumber\\
&\geq&
\sum_{i=1}^m \inf \left\{\left. g_i [V_{\theta}[M]]_{ii}\, \right|\, 
          \mbox{ $M$ is locally unbiased at $\theta$}\right\}\nonumber\\
&=&
\sum_i  g_i \left[J^{S-1}(\theta)\right]^{ii},
\nonumber
\end{eqnarray}
holds true by virtue of the lemma $\ref{lemma:eivei}$,
and that 
the equality in the first inequality does not always establish,
implying
that  in the simultaneous estimation of different parameters,
there is information losses
because of non-commutative nature of the quantum mechanics.

Another proper alternative of the classical Fisher information matrix  is 
 a set 
$\inf {\cal V}_{\theta}({\cal M})$ of symmetric real matrices,
where the notation $\inf$ is defined as follows.
Let us define
\begin{eqnarray}
lb{\cal X}&\equiv&
   \{ A\: |\: \mbox{$A$ is real and symmetric},\,\forall B<A,\,B\in{\cal X}\},
\nonumber\\
ub{\cal X}&\equiv&
   \{ A\: |\: \mbox{$A$ is real and symmetric},\,\forall B>A,\,B\in{\cal X}\},
\nonumber
\end{eqnarray}
where ${\cal X}$ is a set of real symmetric matrices,
and we define $\inf {\cal V}$ by
\begin{eqnarray}
\inf {\cal V}\equiv lb{\cal V}\cap ub(lb{\cal V}).
\nonumber
\end{eqnarray}

Then, we have the following lemma.
\begin{lemma}
$\inf {\cal V}$ is a subset of the boundary $bd{\cal V}$ of ${\cal V}$.
\label{lemma:infbd}
\end{lemma}
This lemma is a straightforward consequence of the following lemma,
which is proved in the appendix B.
\begin{lemma}
If $V$ is an element of ${\cal V}$, then 
$V+V_0$ is also an element of ${\cal V}$,
where $V_0$ is an arbitrary real nonnegative symmetric matrix. 
\label{lemma:v+v0}
\end{lemma}
Because of lemma $\ref{lemma:infbd}$,
it is of interest to determine the boundary $bd {\cal V}$.
$bd {\cal V}$ is turned out to be a subset of ${\cal V}$
such that $V=\CR(G)$ for a weight matrix $G$,
because of lemma $\ref{lemma:v+v0}$, and lemmas 
$\ref{lemma:convex}$-$\ref{lemma:closed}$.
\begin{lemma}
${\cal V}$ is convex.
\label{lemma:convex}
\end{lemma}
\begin{proof}
Let $M$ and $N$ be an unbiased estimator.
Because 
\begin{eqnarray}
\lambda V[M] + (1-\lambda) V[N]=V[\lambda M+(1-\lambda)N]
\nonumber
\end{eqnarray}
holds true and $\lambda M+(1-\lambda)N$ is an unbiased estimator,
we have the lemma.
\end{proof}
\begin{lemma}
${\cal V}$ is closed.
\label{lemma:closed}
\end{lemma}
Lemma $\ref{lemma:closed}$ will be proved 
in the appendix C.

If a model ${\cal M}$ has smaller value of  the attainable CR type bound
at $\theta$
than another model ${\cal N}$ at $\theta'$ has, 
the ${\cal V}_{\theta}({\cal M})$ of 
is located in the `lower part' of $Sym(m)$
compared with that of ${\cal V}_{\theta'}({\cal N})$.

\section{The reduction theorem and the direct approach}
\label{sec:naimark}
\begin{theorem}
(Naimark's theorem, see Ref. $\cite{Holevo:1982}$, pp. 64-68.)\ \\
Any generalized measurement $M$ in ${\cal H}$ can be dilated
to a simple measurement $E$ 
in a larger Hilbert space ${\cal H}'\supset{\cal H}$,
so that
\begin{eqnarray}
M(B)=PE(B)P
\label{eqn:naimark}
\end{eqnarray}
will hold,
where $P$ is the projection from ${\cal H}'$ onto ${\cal H}$.
\end{theorem}
 Naimark's theorem, mixed with commuting theorem,
leads to the following reduction theorem, 
which is essential to our direct approach.
\begin{theorem}
Let  ${\cal M}$ be a $m$-dimensional manifold in ${\cal P}_1$,
and ${\sf B}_{\theta}$ be a system $\{|\phi'\rgl\: |\;|l'_i\rgl,\:i=1,...,m\}$
of vectors in $2m+1$-dimensional Hilbert space ${\cal H}'_{\theta}$
such that
\begin{eqnarray}
\lgl \phi'| l'_j\rgl&=&\lgl \phi| l_j\rgl=0,\nonumber\\
\lgl l'_i| l'_j\rgl&=&\lgl l_i| l_j\rgl,
\nonumber
\end{eqnarray}
for any $i,j$.
Then,
for any locally unbiased estimator $M$ at $\theta$ in ${\cal H}$,
there is a simple measurement $E$ in ${\cal H}'_{\theta}$
such that 
`locally unbiasedness' is satisfied,
\begin{eqnarray}
|x^i\rgl=\int(\hat\theta^i-\theta^i)E(d\theta)|\phi'\rgl\in{\cal H}'_{\theta}
\label{eqn:xe}
\end{eqnarray}
\begin{eqnarray}
\lgl x^i|\phi'\rgl&=&0,\\
{\rm Re}\lgl x^i| l'_j\rgl&=&\delta^i_j
\:(i,j=1,...,m),
\label{eqn:naimark:unbiased}
\end{eqnarray}
and that
the `covariance matrix' $V[E]$ of $E$ equals $V[M]$,
\begin{eqnarray}
V[M]=V[E]\equiv 
\left[\int (\hat\theta^i-\theta^i)(\hat\theta^j-\theta^j)
                                                                                \tr\rho E(d\hat\theta )\right]. 
\label{eqn:ve}                                  
\end{eqnarray}
\label{theorem:simple}
\end{theorem}
\begin{proof}
 For any locally unbiased measurement $M$,
there exists a Hilbert space ${\cal H}_{M}$ and a simple measurement 
$E_{M}$ in ${\cal H}_{M}$ which satisfies $(\ref{eqn:naimark})$ 
by virtue of Naimark's theorem. Note that $E_{M}$ is also locally unbiased.
Let $|y^i\rgl \in{\cal H}_{M}$ denote the estimation vector of $\theta^i$
by $E_M$, that is,
\begin{eqnarray}
|y^i\rgl\equiv\int (\hat\theta^i -\theta^i )E_M(d\hat\theta)|\phi'\rgl.
\nonumber
\end{eqnarray}
Mapping $span_{\bf C}\{|\phi\rgl, |l_i\rgl,|y^i\rgl\: |\:i=1,...,m\}$
isometrically onto ${\cal H}'_{\theta}$ 
so that $\{|\phi\rgl,|l_i\rgl\: |\:i=1...m\}$ 
are mapped to $\{|\phi'\rgl,|l'_i\rgl\: |\: i=1,...,m\}\:$,
we denote the images of $\{|y^i\rgl\: |\:i=1,...,m\}$ by 
$\{|x^i\rgl\: |\: i=1,...,m\}$.

Then, by virtue of the commuting theorem,
we can construct a simple measurement $E$ 
in ${\cal H}'_{\theta}$ satisfying the equations 
$(\ref{eqn:xe})$ - $(\ref{eqn:ve})$.
\end{proof}

The reduction theorem shows that 
${\cal V}$ is identical with the set of matrices
\begin{eqnarray}
V={\rm Re}{\sf X}^*{\sf X}
\nonumber
\end{eqnarray}
such that
\begin{eqnarray}\nonumber
|x^i\rgl\in {\cal H}'_{\theta}\setminus \{|\phi'\rgl\}\:(i=1,...,m),
\label{eqn:xdom}
\end{eqnarray}
where 
${\cal H}'_{\theta}\setminus \{|\phi'\rgl\}$ denotes
the orthogonal complement subspace of ${\cal H}'_{\theta}$,
and that $(\ref{eqn:imxx=0})$ and $(\ref{eqn:naimark:unbiased})$
are satisfied.
Now, the problem is simplified to the large extent,
because 
we only need to treat 
with vectors $\{|x^i\rgl\: |\: i=1,...,m\}$
in finite dimensional Hilbert space  ${\cal H}'_{\theta}$
instead of measurements, or operator valued measures.

We  conclude this  section with 
a corollary of reduction theorem,
which is rather counter-intuitive because
historically, non-projection-valued measurement
is introduced to describe  simultaneous measurements of
non-commuting observables.
\begin{corollary}
When the dimension of ${\cal H}$ is larger than or equal to $2m+1$,
for any unbiased measurement $M$ in ${\cal H}$,
there is a simple measurement $E$ in ${\cal H}$
which has the same covariance matrix as that of $M$.
\label{corollary:ve=vm1}
\end{corollary}
\begin{proof}
Chose
 $\{ |l'_i\rgl\: |\:i=1,...,m\}$
to be $\{ |l_i\rgl\: |\:i=1,...,m\}$.
\end{proof}
Especially, if  ${\cal H}$ is infinite dimensional,
as is the space of wave functions,
the assumption of the corollary is always satisfied.

\section{Lagrange's method of indeterminate 
coefficients in the pure state estimation theory}
\label{sec:lagrange}
Now, we apply our direct approach to the problem presented in 
the section $\ref{sec:nonclassical}$, or 
the minimization of the functional 
$\Tr G{\rm Re}{\sf X}^*{\sf X}$ of vectors in ${\cal H}'_{\theta}$.
One of most straightforward approaches to this problem is
 Langrange's indeterminate coefficients method.
First,
denoting an ordered pair $\{|l_i'\: |\: i=1,...,\}$
of vectors in ${\cal H}'_{\theta}$ also by ${\sf L}$,
the symbol which is used also for 
an ordered pair $\{|l_i\: |\: i=1,...,\}$
of vectors in ${\cal H}$,
we define a function $Lag({\sf X})$ by 
\begin{eqnarray}
Lag({\sf X})\equiv
{\rm Re}{\Tr}{\sf X}^*{\sf X}G
-2{\Tr}(({\rm Re}{\sf X}^*{\sf L}-I_m)\Xi)
-{\Tr}{\rm Im}{\sf X}^*{\sf X}\Lambda,
\label{eqn:lagrangean}
\end{eqnarray}
where $\Xi, \Lambda$ are real matrices whose components are 
Langrange's indeterminate coefficients.
Here, $\Lambda$ can be chosen to be antisymmetric, for
\begin{eqnarray} 
{\Tr}{\rm Im}{\sf X}^*{\sf X}\Lambda
={\Tr}{\rm Im}{\sf X}^*{\sf X}(\Lambda-\Lambda^T)/2
\nonumber
\end{eqnarray}
holds true and only antisymmetric part of $\Lambda$ appears
in $(\ref{eqn:lagrangean})$.

From here, we follow the routine of  
Langrange's method of indeterminate coefficients.
Differentiating $L({\sf X}+\epsilon\delta{\sf X})$ with respect to $\epsilon$
and substituting  $0$ into $\epsilon$ in the derivative,
we get
\begin{eqnarray}
{\rm Re}{\Tr}(\delta{\sf X}^*
(2{\sf X}G-2{\sf L}\Xi-2i{\sf X}\Lambda))=0.
\nonumber
\end{eqnarray}
Because $\delta{\sf X}$ is arbitrary,
\begin{eqnarray}
{\sf X}(G-i\Lambda)={\sf L}\Xi
\label{eqn:xg-il}
\end{eqnarray}
is induced. 

Multipling ${\sf X}^*$ to both sides of 
$(\ref{eqn:xg-il})$,
the real part of the outcomming equation, 
together with $(\ref{eqn:lagrange:restriction1})$, yields
\begin{eqnarray}
\Xi={\rm Re}{\sf X}^*{\sf X}G=VG.
\label{eqn:lagrange:xi}
\end{eqnarray}
Substituting $(\ref{eqn:lagrange:xi})$ into $(\ref{eqn:xg-il})$ ,
we obtain
\begin{eqnarray}
{\sf X}(G-i\Lambda)={\sf L}VG.
\label{eqn:basic0.1}
\end{eqnarray}

In this paper,
we solve 
$(\ref{eqn:lagrange:restriction1})$,
$(\ref{eqn:imxx=0})$,
$(\ref{eqn:basic0.1})$ and $V={\rm Re}{\sf X}^*{\sf X}$
with respect to ${\sf X}$, real symmetric matrix $V$ and 
real antisymmetric matrix $ \Lambda$,
for the variety of pure state models.
However,
the general solution is still far out of our reach.

\section{The model with two parameters}
\label{sec:2parameter}
In this section, we determine the boundary of the set ${\cal V}$
in the case of the $2$-dimensional model.

The equation$(\ref{eqn:basic0.1})$, 
mixed  with $(\ref{eqn:imxx=0})$,  leads to
\begin{eqnarray}
(G-i\Lambda)V(G-i\Lambda)
=GV{\sf L}^*{\sf L}VG.
\label{eqn:basic1.0}
\end{eqnarray}
whose 
real part and imaginary part  are
\begin{eqnarray}
GVG-\Lambda V \Lambda =GV J^S VG,
\label{eqn:basic1.1}
\end{eqnarray}
and
\begin{eqnarray}
GV\Lambda +\Lambda VG=-GV\tilde{J}VG,
\label{eqn:basic1.2}
\end{eqnarray}
where $\tilde J$ denotes ${\rm Im}{\sf L}^*{\sf L}$,
respectively.

As is proved in the following,
when the matrix $G$ is strictly positive, 
$(\ref{eqn:basic1.0})$ is equivalent to the existence of ${\sf X}$
which satisfies
$(\ref{eqn:lagrange:restriction1})$,
$(\ref{eqn:imxx=0})$,
$(\ref{eqn:basic1.0})$,
and $V={\rm Re}{\sf X}^*{\sf X}$.
If $V$ and $\Lambda$ satisfying $(\ref{eqn:basic1.0})$ exist,
${\sf X}$ which satisfies  $(\ref{eqn:basic0.1})$ and 
$(\ref{eqn:imxx=0})$ is given by 
${\sf X}=UV^{1/2}$, where $U$ is such a $2m+1\times m$  complex matrix that
$U^*U=I_m$.
${\sf X}=UV^{1/2}$ also satisfies
$(\ref{eqn:lagrange:restriction1})$,
because 
\begin{eqnarray}
VG={\rm Re}{\sf X}^*{\sf L}VG\nonumber
\end{eqnarray}
 is obtained by
multipling ${\sf X}^*$ to 
and taking real part of the both sides of $(\ref{eqn:basic0.1})$.

Hence, if $G$ is strictly positive, 
our task is to solve $(\ref{eqn:basic1.1})$ and  $(\ref{eqn:basic1.2})$
for real positive symmetric matrix $V$ and real antisymmetric matrix $\Lambda$.
When $G$ is not strictly positive, 
after solving $(\ref{eqn:basic1.1})$ and  $(\ref{eqn:basic1.2})$,
we must check 
whether there exists such ${\sf X}$ which satisfies
$(\ref{eqn:lagrange:restriction1})$, $(\ref{eqn:imxx=0})$
and 
$V={\rm Re}{\sf X}^*{\sf X}$.

Throughout  this section, we parameterize the model so that
$J^S$ is equal to the identity matrix $I_m$.
Given an arbitrary coordinate system $\{\theta^i|\: i=1,...,m\}$,
such a coordinate system $\{\theta'^i|\: i=1,...,m\}$
is obtained by the following coordinate transform:
\begin{eqnarray}
\theta'^i=\sum_{j=1}^m [(J^{S})^{1/2}]_{ij}\:\theta^j\:(i=1,...,m).
\label{eqn:2para:trans1}
\end{eqnarray}
By this coordinate transform,
$V$ is transformed as: 
\begin{eqnarray}
V'&=&(J^{S})^{1/2} V (J^{S})^{1/2}.
\label{eqn:2para:trans2}
\end{eqnarray}
If the result in the originally given coordinate is needed,
one  only needs to transform the result 
in the coordinate system $\{\theta'^i|\: i=1,...,m\}$
using $(\ref{eqn:2para:trans2})$ in the converse way.

So far, we have not assumed $\dim{\cal M}=2$.
When $\dim{\cal M}=2$, covariance matrices are included
in the space $Sym(2)$ of  $2\times 2$ symmetric matrices
which is parameterized by $x,y,$ and $z$, where
\begin{eqnarray}
Sym(2)=
\left\{V\left|V=
\left[
\begin{array}{cc}
z+x & y\\
y   & z-x
\end{array}
\right]
\right.
\right\}.
\nonumber
\end{eqnarray}
Before tackling the equations $(\ref{eqn:basic1.1})$ 
and  $(\ref{eqn:basic1.2})$, three useful facts about this parameterization
are noted.
First, letting $A$ is a symmetric real matrix
which is represented by $(A_x, A_y, A_z)$ in the $(x,y,z)$-space,
the set ${\cal C}_+(A)$ of all matrices larger than $A$
is
\begin{eqnarray}
{\cal C}_+(A)
=\{(x,y,z)\:|\: (z-A_z)^2-(x-A_x)^2-(y-A_y)^2\ge 0,\: z\geq A_z\},
\nonumber
\end{eqnarray}
that is, inside of a upside-down corn with its vertex at $A=(A_x, A_y, A_z)$.
Hence,
$\cal V$ is a subset of ${\cal C}_+(I_m)$, or inside of 
a upside-down corn with its vertex 
at $(0,0,1)$ because of the SLD CR inequality. 
When the model $\cal M$ is classical at $\theta$,
$\cal V$ coincides with  ${\cal C}_+(I_m)$.

Second, an  action of 
rotation matrix $R_{\theta}$ to $V$ such that 
$R_{\theta}V R_{\theta}^T$,
where 
\begin{eqnarray}
R_{\theta}=\left[
\begin{array}{cc}
\cos\theta & -\sin\theta\\
\sin\theta & \cos\theta
\end{array}
\right],\nonumber
\end{eqnarray}
corresponds to the rotation in the $(x,y,z)$-space around 
$z$-axis by the angle $2\theta$.

Third, we have the following lemma.
\begin{lemma}
${\cal V}$ is rotationally symmetric around $z$-axis,
if $\cal M$ is parameterized so that
$J^S$ writes the unit matrix $I_m$.
\end{lemma}
\begin{proof}
The necessary and sufficient condition 
for $\cal V$ to have rotational symmetry around $z$-axis
is 
the existence of  a $2m+1$ by $m$ complex matrix ${\sf Y}$ satisfying 
$(\ref{eqn:lagrange:restriction1})$, $(\ref{eqn:imxx=0})$
and
\begin{eqnarray}
\forall\theta\;\; 
R_{\theta}\,({\rm Re}{\sf X}^*{\sf X})\,R_{\theta}^T
=R_{\theta}{\sf X}^*{\sf X}R_{\theta}^T
={\sf Y}^*{\sf Y},
\label{eqn:rxxr}
\end{eqnarray}
for any given ordered pair ${\sf X}$ 
of vectors which satisfies
$(\ref{eqn:lagrange:restriction1})$ and $(\ref{eqn:imxx=0})$.

On the other hand, because of ${\sf L}^*{\sf L}=I_m+i\tilde{J}$,
elementary calculation shows
\begin{eqnarray}
{\sf L}^*{\sf L}=R_{\theta}{\sf L}^*{\sf L}R_{\theta}^T,
\nonumber
\end{eqnarray}
or equivalently, for some unitary transform in 
${\cal H}'_{\theta}\setminus\{|\phi\rgl\}$,
\[
{\sf L}^*U=R_{\theta}{\sf L}^*,
\]
which leads, together with $(\ref{eqn:lagrange:restriction1})$, to
\begin{eqnarray}
{\rm Re}{\sf L}^*U{\sf X}=R_{\theta}.
\nonumber
\end{eqnarray}

Therefore,
\[
{\sf Y}=U{\sf X}R_{\theta}^T
\]
satisfies $(\ref{eqn:rxxr})$, and we have the lemma.
\end{proof}

Now,
the boundary of the intersection $\tilde{\cal V}$ of $\cal V$ and $zx$-plain
is to be calculated,
because $\cal V$ is obtained by rotating $\tilde{\cal V}$ around $z$-axis,
by virtue of this lemma.
$bd\tilde{\cal V}$ is obtained as the totality of
the matrix $V={\rm Re}{\sf X}^*{\sf X}$ which satisfies
$(\ref{eqn:lagrange:restriction1})$, $(\ref{eqn:imxx=0})$,
and $(\ref{eqn:basic0.1})$,
for a diagonal real nonnegative matrix $G$.

Let us begin with the case where a diagonal matrix $G$
is positive definite.
In this case, we only need to deal with 
$(\ref{eqn:basic1.1})$ and $(\ref{eqn:basic1.2})$.
Let
\begin{eqnarray}
\tilde{J}=\left[
\begin{array}{cc}
0 & -\beta\\
\beta & 0
\end{array}
\right],
\nonumber
\end{eqnarray}
and 
\begin{eqnarray}
\Lambda=\left[
\begin{array}{cc}
0 & -\lambda\\
\lambda & 0
\end{array}
\right],\:
G=\left[
\begin{array}{cc}
1 & 0 \\
0 & g
\end{array}
\right],\:
V=\left[
\begin{array}{cc}
u & 0 \\
0 & v
\end{array}
\right],
\nonumber
\end{eqnarray}
where $g,v,$ and $u$ are positive real real numbers.
Note that
\begin{eqnarray} 
|\beta|\leq 1
\nonumber
\end{eqnarray}
holds, because ${\sf L}^*{\sf L}=I_m+\tilde{J}$
is nonnegative definite.
Then,  $(\ref{eqn:basic1.1})$ and $(\ref{eqn:basic1.2})$ writes
\begin{eqnarray}
u+v\lambda^2-u^2=0,\nonumber\\
vg^2+u\lambda^2-v^2g^2=0,\nonumber\\
vg\lambda+u\lambda+uv\beta g=0.
\label{eqn:2para:uvglmd}
\end{eqnarray}
The necessary and sufficient condition for 
$\lambda$ and positive $g$ to exist is,
after some calculations,
\begin{eqnarray}
\sqrt{u-1}+\sqrt{v-1}-|\beta|\sqrt{uv}=0.
\label{eqn:2para:uv}
\end{eqnarray}
Note that  $u$ and $v$ are larger than  or equal to $1$, 
because $V\ge J^{S-1}=I_m$.
Substitution of  $u=z+x$ and $v=z-x$ into $(\ref{eqn:2para:uv})$ and 
some calculation leads to
\begin{eqnarray}
|\beta|\sqrt{(z+x-1)(z-x-1)}
\pm\sqrt{1-\beta^2}\left(\sqrt{z+x-1}+\sqrt{z-x-1}\right)=|\beta|.
\nonumber
\end{eqnarray} 
It is easily shown that the lower sign in the equation 
corresponds to  the set of stationary points, and
\begin{eqnarray}
|\beta|\sqrt{(z+x-1)(z-x-1)}
\pm\sqrt{1-\beta^2}\left(\sqrt{z+x-1}+\sqrt{z-x-1}\right)=|\beta|
\nonumber\\
\label{eqn:2para:zx1}
\end{eqnarray}
gives a part of $bd\tilde{\cal V}$.
In $(\ref{eqn:2para:zx1})$, $x$ takes value ranging 
from $-\beta^2/(1-\beta^2)$ to  $\beta^2/(1-\beta^2)$ if $|\beta|$ is smaller 
than $1$.
When $|\beta|=1$, $x$ varies from $-\infty$ to $\infty$.
This restriction on the range of $x$ comes from
the positivity of $z-x-1$ and $z+x-1$.

\begin{figure}
\begin{center}
\epsfbox{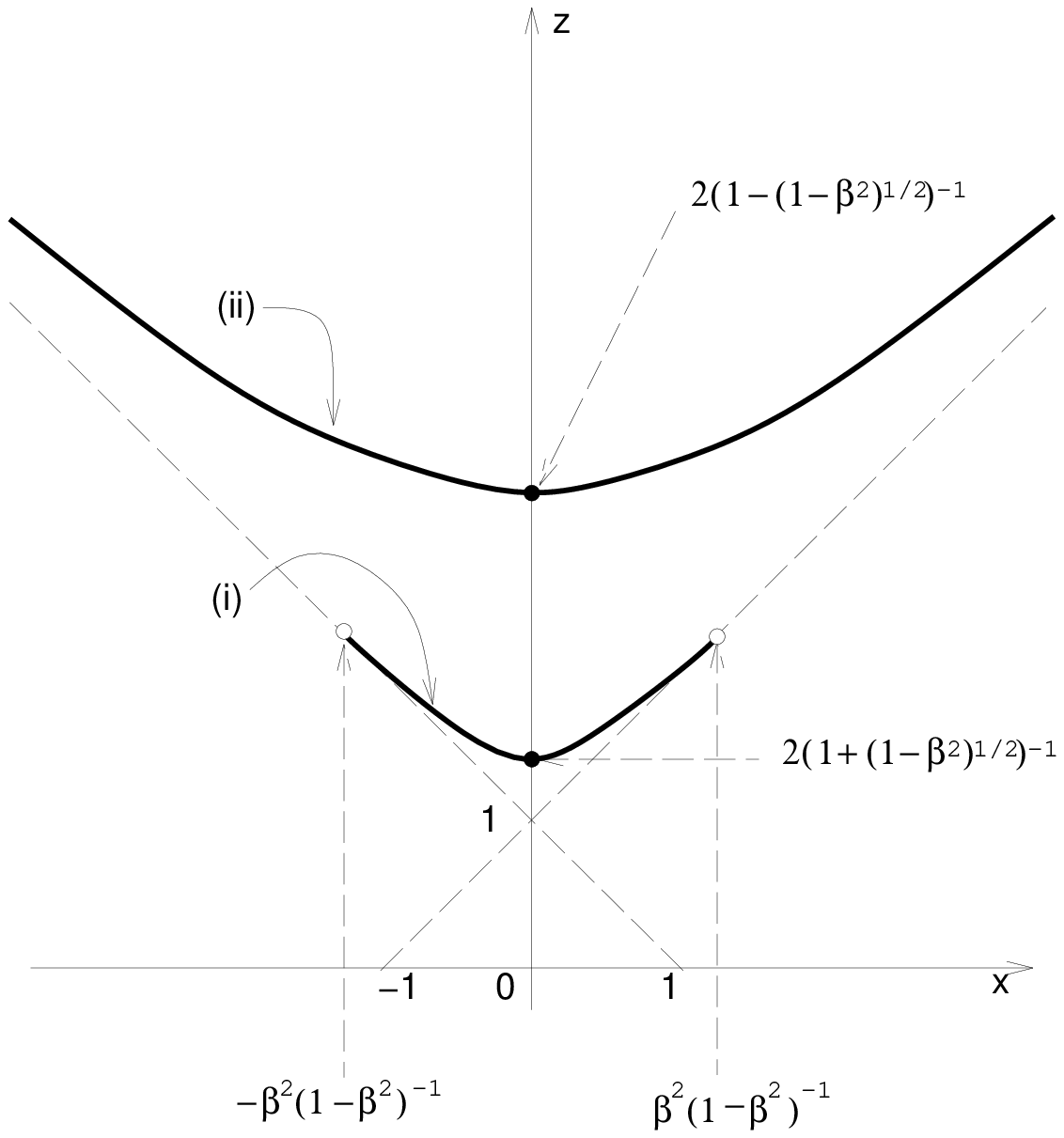}
\caption[Two stationary lines]{
Two stationary lines;\\
$(i)\: |\beta|\sqrt{(z+x-1)(z-x-1)}
+\sqrt{1-\beta^2}\left(\sqrt{z+x-1}+\sqrt{z-x-1}\right)=|\beta|;$\\
$(ii)\: |\beta|\sqrt{(z+x-1)(z-x-1)}
-\sqrt{1-\beta^2}\left(\sqrt{z+x-1}+\sqrt{z-x-1}\right)=|\beta|;$
}
\label{fig:2p:fig1}
\end{center}
\end{figure}

When 
\begin{eqnarray}
G=\left[
\begin{array}{cc}
1 & 0\\
0 & 0
\end{array}
\right],\:\:\mbox{or}\:\:
G=\left[
\begin{array}{cc}
0 & 0\\
0 & 1
\end{array}
\right],
\label{eqn:2para:G}
\end{eqnarray}
we must treat the case of  $|\beta|=1$ and
the case of $|\beta|<1$ differently.
In the case of $|\beta|=1$, 
there exists no  $2m \times m $ complex matrix ${\sf X}$
which satisfies  $V={\rm Re}{\sf X}^*{\sf X}$, 
$(\ref{eqn:lagrange:restriction1})$, $(\ref{eqn:imxx=0})$ and 
$(\ref{eqn:basic1.0})$.
On the other hand, if $|\beta|<1$,
such complex matrix ${\sf X}$ always exists and 
$V={\rm Re}{\sf X}^*{\sf X}$ is given by, in terms of $(x,y,z)$,
\begin{eqnarray}
z&=&-x+1,\: x\leq -\frac{\beta^2 }{1-\beta^2}\nonumber\\
&&\mbox{or}\nonumber\\
z&=&x+1,\: x\ge \frac{\beta^2 }{1-\beta^2}
\label{eqn:2para:zx2}
\end{eqnarray}

Because any element on the line $(\ref{eqn:2para:zx2})$,
if $x\neq \pm\beta^2/(1-\beta^2)$, has
an element of ${\cal V}$ which is smaller than itsself,
$(\ref{eqn:2para:zx1})$
 the intersection of $\inf{\cal V}$ and $zx$-plane,
where
\begin{eqnarray}
-\beta^2/(1-\beta^2)\leq x \leq \beta^2/(1-\beta^2).
\nonumber
\end{eqnarray}
The intersection of $z$-axis and $bd{\cal V}$ gives
\begin{eqnarray}
\CR(J^S)=\frac{4}{1+\sqrt{1-|\beta|^2}},
\label{eqn:minvv}
\end{eqnarray}
where the equality holds in any parameterization of the model ${\cal M}$.

In arbitrary parameterization of the model ${\cal M}$,
with help of $(\ref{eqn:2para:zx1})$ and $(\ref{eqn:2para:trans2})$, 
$\inf({\cal V})$ is obtained as, for $|\beta|> 0$,
\begin{eqnarray}
&&\det \sqrt{\tilde{V}(V)}
   +\sqrt{(1/\beta^2)-1}\,\Tr\sqrt{\tilde{V}(V)}=1\: (|\beta|> 0),
\nonumber\\
&&\Tr\sqrt{\tilde{V}(V)}=0\: (\beta=0),
\label{eqn:exv}
\end{eqnarray}
where
\begin{eqnarray}
\tilde{V}(V)\equiv\sqrt{J^S}V\sqrt{J^S}-I_m.
\nonumber
\end{eqnarray}

Slight look at the equations $(\ref{eqn:exv})$ 
leads to the  following theorem.
\begin{theorem}
In the 2-dimensional model,
if
\begin{eqnarray}
&&|\beta(\theta,{\cal M})|\geq |\beta(\theta',{\cal M}')|,\nonumber\\
&&J^S(\theta,{\cal M})=J^S(\theta',{\cal M}'),
\nonumber
\end{eqnarray}
then
the ${\cal V}_{\theta}({\cal M})$
is a subset of ${\cal V}_{\theta'}({\cal M}')$.
\label{theorem:2pcalv}
\end{theorem}
The equations $(\ref{eqn:exv})$  and tedious but elementary calculations
shows the following theorem.
\begin{theorem}
In the 2-dimensional model, if
\begin{eqnarray}
&&|\beta(\theta,{\cal M})|= |\beta(\theta',{\cal M}')|,\nonumber\\
&&J^S(\theta,{\cal M})\leq J^S(\theta',{\cal M}'),
\nonumber
\end{eqnarray}
then
the ${\cal V}_{\theta}({\cal M})$
is a subset of ${\cal V}_{\theta'}({\cal M}')$.
\end{theorem}
By virtue of these theorems, 
$|\beta|$ can be seen as a measure of `uncertainty'
between the two parameters.
Two extreme cases are worthy of special attention;
When $|\beta|=0$, the model $\cal M$ is classical at $\theta$
and ${\cal V}$ is maximum.
On the other hand,
if $|\beta|=1$ , ${\cal V}$ is minimum  and 
uncertainty between  $\theta^1$ and $\theta^2$ is maximum.
In the latter case, we say that the model is {\it coherent} at $\theta$. 

\begin{figure}[htp]
\begin{center}
\leavevmode\epsfxsize=6.5cm\epsfbox{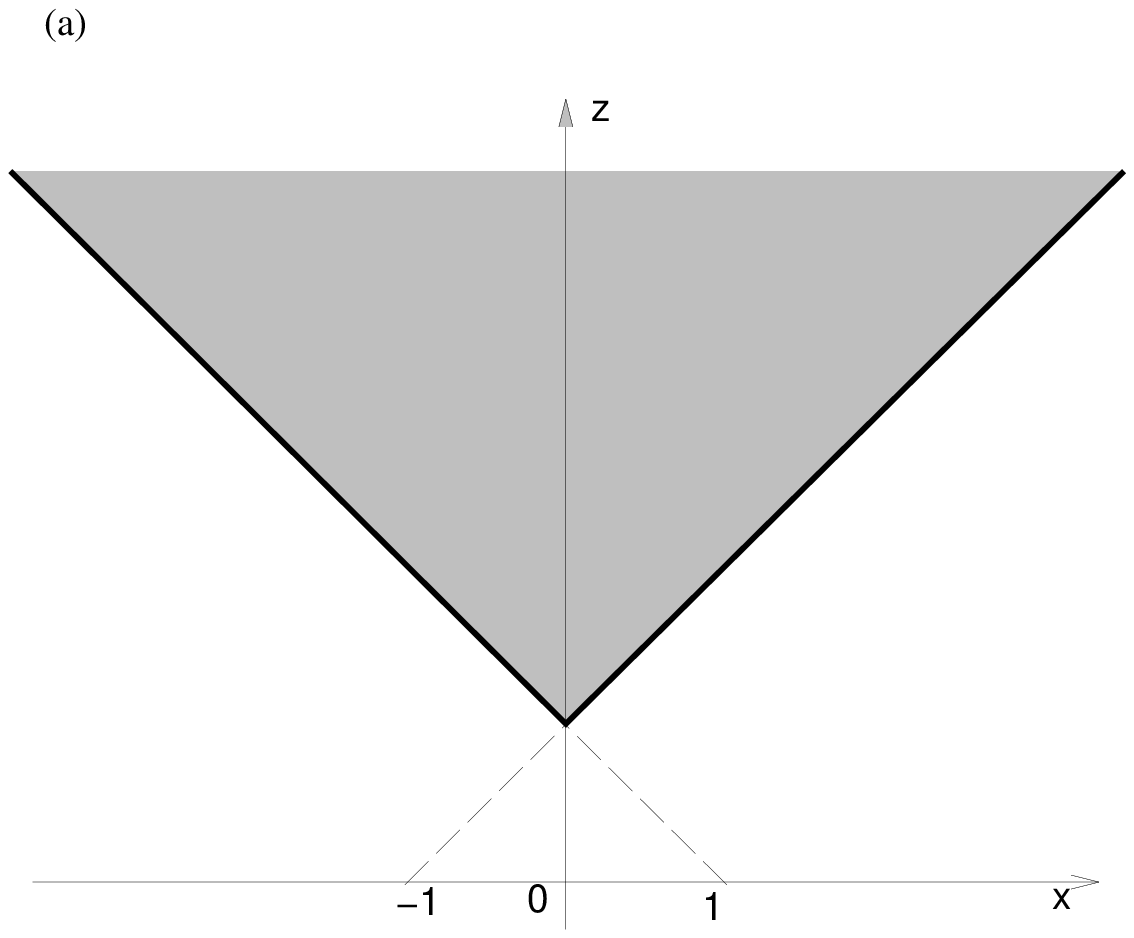}\\
\leavevmode\epsfxsize=6.5cm\epsfbox{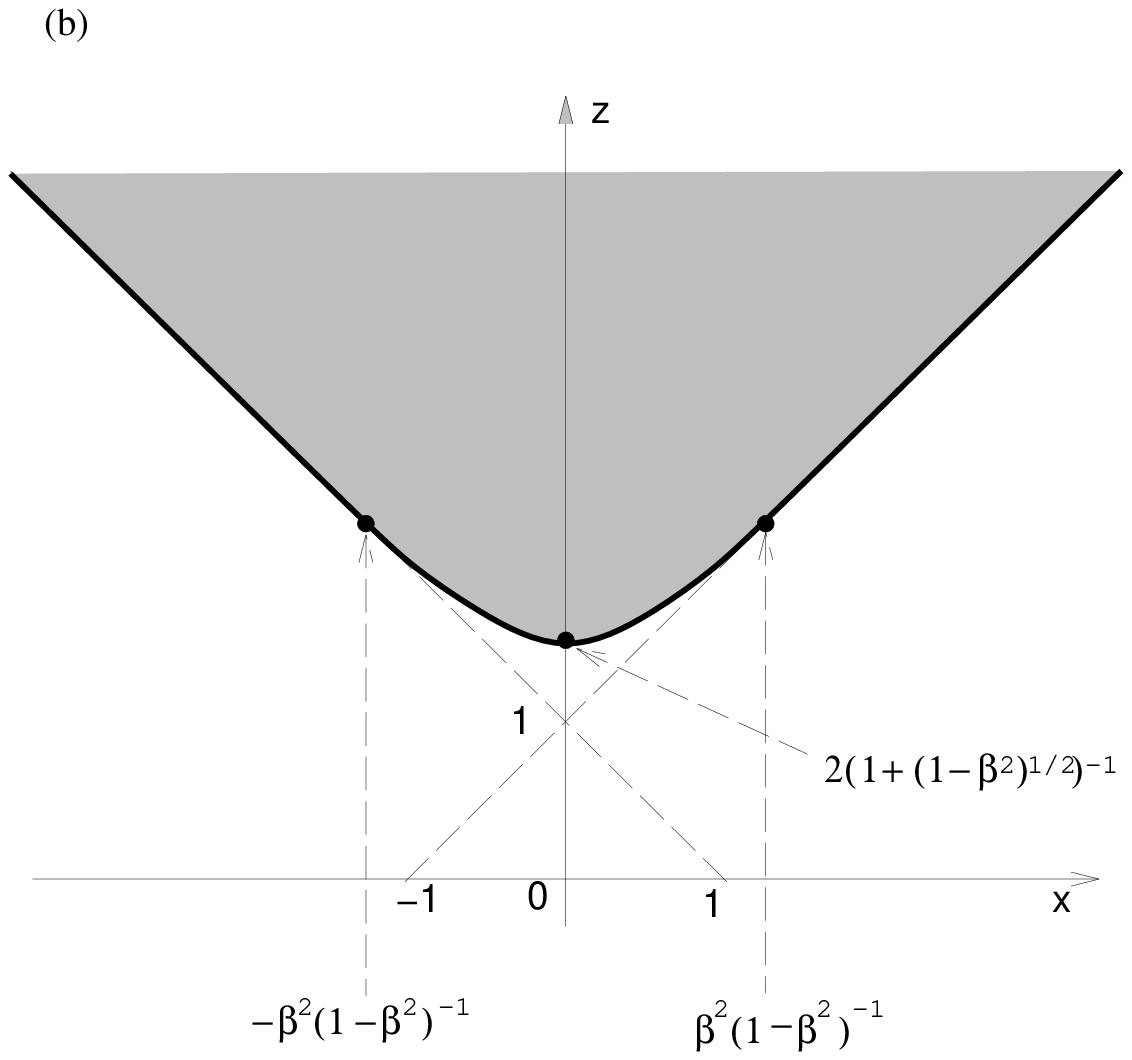}\\
\leavevmode\epsfxsize=6.5cm\epsfbox{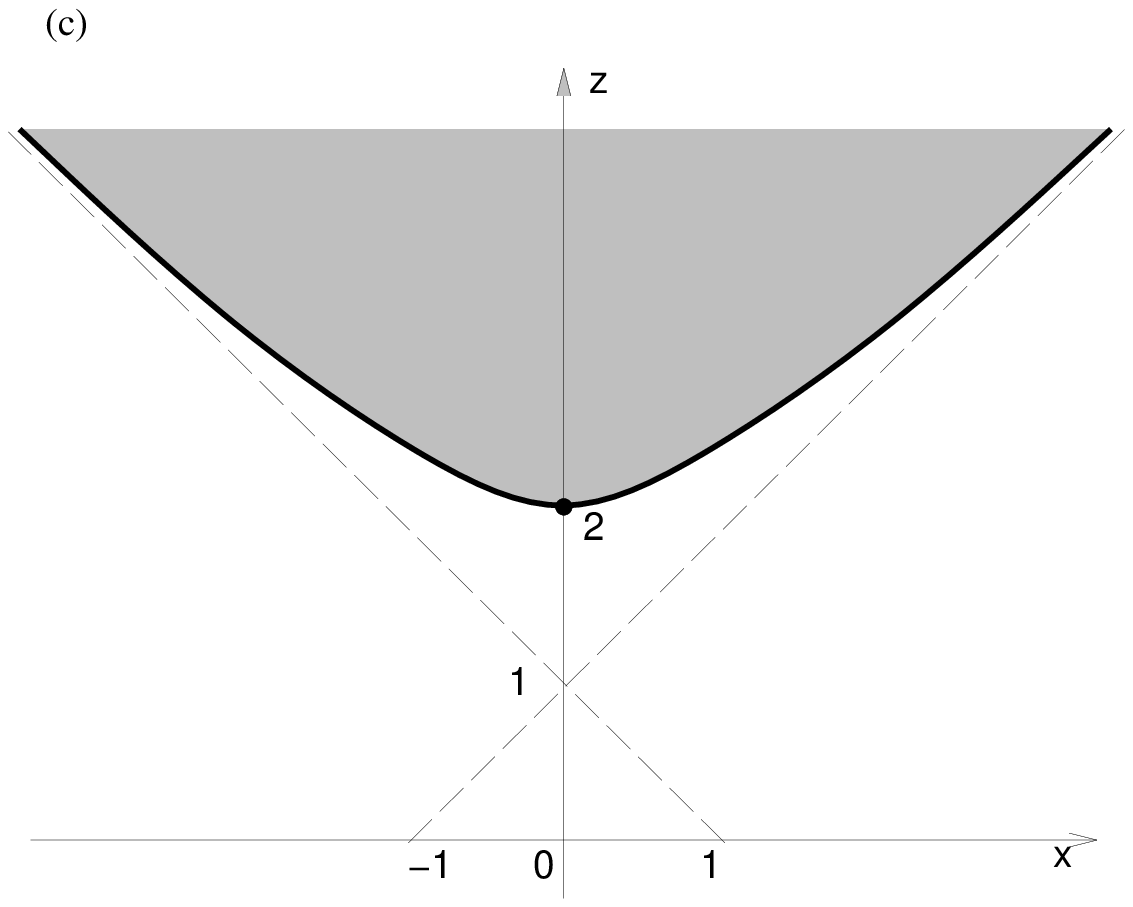}
\caption{(a)$\;|\beta|=0;\:$(b)$\;0<|\beta|<1;\;$(c)$\;|\beta|=1$.}
\label{c1}
\end{center}
\end{figure}

\pagebreak

\begin{example}(spin rotation model)
We define {\it spin rotation model} \cite{Abe} by
\begin{eqnarray}
&{\cal M}_{s,m}&=\pi({\cal N}),\nonumber\\
&{\cal N}_{s,m}&= \left\{ |\phi(\theta)\rgl\, \left|\, 
        |\phi(\theta)\rgl=T(\theta)|s, m\rgl,\,
        0\leq\theta^1<\pi,\,0\leq\theta^2<2\pi\right. \right\},\nonumber\\
&& T(\theta)=\exp \left(i\theta^1(\sin\theta^2 S_x-\cos\theta^2 S_y\right),
\label{eqn:gscoherent}
\end{eqnarray}
where $S_x$, $S_y$, $S_z$ are spin operators,
and $|s,m\rgl$ is defined by,
\begin{eqnarray}
&&S_z|j,m\rgl=\hbar m |s, m\rgl,\nonumber\\
&&\left(S^2_x+S^2_y+S^2_z\right)|s,m\rgl=\hbar^2 s(s+1)|s, m\rgl.
\nonumber
\end{eqnarray}
$s$ takes value of half integers,
and m is a half integer such that $-j\leq m\leq j$.
Then after tedious calculations, we obtain 
\begin{eqnarray}
\beta(\theta, {\cal M}_{s,m}) =\frac{m}{s^2+ s -m^2}.
\nonumber
\end{eqnarray}
If $m=\alpha s$, where $\alpha<1$ is a constant,
$\beta(\theta, {\cal M}_{s,m})$ tends to zero as $s\rightarrow\infty$,
and the model ${\cal M}_{s,m}$ becomes  quasi-classical.
However, if  $m=s$, the model ${\cal M}_{s,m}$ is coherent for any $s$.
\end{example}
\begin{example}(shifted number state model)
{\it shifted number state model}, which has four parameters, is defined by
\begin{eqnarray}
{\cal M}_n &=&\pi({\cal N}_n),\nonumber\\
{\cal N}_n &=&
\left\{|\phi(\theta)\rgl\: \left|\; 
     |\phi(\theta)\rgl=D(\theta)|n\rgl,\; \theta\in\R^2\right.\right\},
\nonumber
\end{eqnarray}
where letting 
$P$, $X$ be  the momentum operator and the position operator respectively,
\begin{eqnarray}
D(\theta)&\equiv&\exp\frac{i}{\hbar}\left(-\theta^1 X+ \theta^2 P\right),
\nonumber
\end{eqnarray}
and $|n\rgl$ is the $n$th eigenstate of the harmonic oscillator,
\begin{eqnarray}
H=-\frac{1}{2} P^2+ \frac{1}{2}X^2.
\nonumber
\end{eqnarray}
After some calculations, we have
\begin{eqnarray}
\beta(\theta, {\cal M}_n)=\frac{-1}{2n+1}.
\nonumber
\end{eqnarray}
As $n$ tends to infinity, $\beta(\theta, {\cal M}_n)$ goes to 0 and
the model becomes quasi-classical.
\end{example}

\section{Informational exclusiveness and independence, and 
         direct sum of the models}
In a $m$-dimensional model ${\cal M}$,
we say parameter $\theta^i$ and $\theta^j$ are 
{\it informationally independent} at $\theta_0$, iff
\begin{eqnarray}
{\rm Re}\lgl l_i|l_j\rgl |_{\theta=\theta_0}
={\rm Im}\lgl l_i|l_j\rgl |_{\theta=\theta_0}
=0,
\nonumber
\end{eqnarray}
because, if the equation holds true,
letting the submodels 
${\cal M}(1|\theta_0)$, ${\cal M}(2|\theta_0)$ 
and ${\cal M}(1,2|\theta_0)$ of ${\cal M}$,  be
\begin{eqnarray}
{\cal M}(1|\theta_0)&\equiv&
\{\rho(\theta)\:|\: \theta=(\theta^1,\theta^2_0,...,\theta^m_0),\,
\theta^1\in\R\},
\nonumber\\
{\cal M}(2|\theta_0)&\equiv&
\{\rho(\theta)\:|\: \theta=(\theta^1_0,\theta^2,...,\theta^m_0),\,
\theta^2\in\R\},
\nonumber\\
{\cal M}(1,2|\theta_0)&\equiv&
\left\{\rho(\theta)\:\left|\: 
\theta=(\,\theta^1,\,\theta^2,\,\theta^3_0 ,...,\,\theta^m_0\,),\,
(\theta^1,\,\theta^2\,)\in \R^2\right.\right\},
\label{eqn:calm12}
\end{eqnarray}
the following equality establishes:
\begin{eqnarray}
\CR(\theta_0, diag(g_1, g_2),{\cal M}(1,2|\theta_0))
= \CR(\theta_0, diag(g_1),{\cal M}(1|\theta_0))
  +\CR(\theta_0, diag(g_2),{\cal M}(2|\theta_0)),
\nonumber
\end{eqnarray}
which means that
in the simultaneous estimation of the parameter $(\theta^1,\theta^2)$,
both of the parameters can be estimated without the loss of information
compared with the estimation of each parameters.

On the other hand, iff
\begin{eqnarray}
{\rm Re}\lgl l_1|l_2\rgl |_{\theta=\theta_0}=0,
\label{eqn:def:excv}
\end{eqnarray}
and ${\cal M}(1,2|\theta_0)$ is coherent, or equivalently,
\begin{eqnarray}
{\rm Im}\lgl l_1|l_2\rgl |_{\theta=\theta_0}
=\left.(\lgl l_1|l_1\rgl\lgl l_2|l_2\rgl)^{1/2}\right|_{\theta=\theta_0}
\nonumber
\end{eqnarray} 
hold true,
we say the parameters are {\it informationally exclusive} at $\theta_0$,
because of the following theorem.
\begin{theorem}
Let  $\theta^1$ and  $\theta^2$ be 
informationally exclusive parameters at $\theta_0$,
and  $M'$ a measurement which takes value in $\R^2$
and satisfies local unbiasedness condition about $\theta^1$ at $\theta_0$,
\begin{eqnarray}
\int\hat\theta^1\Tr\rho(\theta_0)M'(d\hat\theta)=\theta^1_0,\nonumber\\
\int\hat\theta^1\Tr\left.\frac{\partial \rho}{\partial \theta^2}
\right|_{\theta=\theta_0}M'(d\hat\theta)=0
\label{eqn:th1unb}
\end{eqnarray}
If the  measurement $M'$ satisfies
{\it i.e.},
\begin{eqnarray}
\int(\hat\theta^1-\theta^1_0)^2\Tr\rho(\theta_0)M'(d\hat\theta)
=\CR(diag(1,0),\theta_0)=\left[J^{S-1}\right]^{11},
\label{eqn:excv:1}
\end{eqnarray}
$M'$ can extract no information about
$\theta^2$ from the system, {\it i.e.},
\begin{eqnarray}
\forall B\subset {\R}^2 \:\:\:
\Tr \left(M'(B)
\left. \frac{\partial \rho}{\partial \theta^2}
\right|_{\theta=\theta_0}
\right)=0,
\label{eqn:excv:2}
\end{eqnarray}
and vice versa.
\label{theorem:exclusive}
\end{theorem}
\begin{proof}
Let $E$ be a Naimark's dilation of the measurement $M'$, and 
decompose the estimation vector 
$|x\rgl\equiv|x[E,|\phi(\theta_0)\rgl]\rgl$  of $E$ as
\begin{eqnarray}
|x\rgl=z|\phi(\theta_0)\rgl+w|l_1(\theta_0)\rgl+|\psi\rgl,
\nonumber
\end{eqnarray}
where $|\psi\rgl$ is orthogonal to both of $|\phi(\theta_0)\rgl$
and $|l_1(\theta_0)\rgl$.
Then, local unbiasedness condition $(\ref{eqn:th1unb})$
leads to $z=0$ and $w=\left[J^{S-1}\right]^{11}$.
$|\psi\rgl$ must be the zero vector
for  $M'$ to achieve the the equality $(\ref{eqn:excv:1})$,
because the variance of $M'$ writes
 \begin{eqnarray}
\int(\hat\theta^1-\theta^1_0)^2\Tr\rho(\theta_0)M'(d\hat\theta)
=\lgl x|x \rgl=\left[J^{S-1}\right]^{11}+\lgl \psi|\psi\rgl.
\nonumber
\end{eqnarray}

Using the fact  that  by virtue of informational exclusiveness,
$|l_2(\theta_0)\rgl$
writes 
\begin{eqnarray}
|l_2(\theta_0)\rgl= ia|l_1(\theta_0)\rgl,
\nonumber
\end{eqnarray}
where $i$ is the imaginary unit and $a$ a real number,
 we can check the equality $(\ref{eqn:excv:2})$
by the following calculations:
\begin{eqnarray}
& &\Tr \left( M'(B)
   \left. \frac{\partial \rho}{\partial \theta^2}
   \right|_{\theta=\theta_0}
    \right)\nonumber\\
&=&{\rm Re}\,\lgl \phi(\theta_0)| E(B) |l_2\rgl \nonumber\\
&=&-a{\rm Im}\,\lgl \phi(\theta_0)| E(B) |l_1\rgl \nonumber\\
&=&-a\left[J^{S}\right]_{11}
 {\rm Im}\,\lgl \phi(\theta_0)| E(B)
 \int(\hat\theta^1-\theta^1_0)E(d\hat\theta)|\phi(\theta_0)\rgl \nonumber\\
&=&0.\nonumber
\end{eqnarray}
\end{proof}

Fujiwara and Nagaoka $\cite{FujiwaraNagaoka:1996}$
showed that in the $2$-dimensional model
with the informationally exclusive parameters,
 the best strategy for the estimation is 
 alternative 
application of the best measurement for each parameter to the system.
This fact is quite natural in the light of 
theorem $\ref{theorem:exclusive}$.

For the submodels 
\begin{eqnarray}
{\cal M}_1\equiv{\cal M}(1,2,...,m_1\,|\,\theta_0),\:
{\cal M}_2\equiv{\cal M}(m_1,m_1+1,...,m\,|\,\theta_0) 
\nonumber
\end{eqnarray}
of ${\cal M}$,
which are defined almost 
in the same way as the definition $(\ref{eqn:calm12})$ of
${\cal M}(1,2\,|\,\theta_0)$,
we say that 
${\cal M}$ is the sum of ${\cal M}$ and ${\cal M}$ at $\theta_0$,
and express the fact by the notation,
\begin{eqnarray}
{\cal M}|_{\theta_0}={\cal M}_1\oplus {\cal M}_2|_{\theta_0}.
\nonumber
\end{eqnarray}
Throughout the section, $m-m_1$ is denoted by $m_2$.
\begin{lemma}
If any parameter of ${\cal M}_1$ is informationally independent
of any parameter of ${\cal M}_2$ at $\theta_0$, and
the weight matrix $G$ writes
\begin{eqnarray}
G=\left[
\begin{array}{cc}
G_1 & 0\\
0 & G_2
\end{array}
\right],
\nonumber
\end{eqnarray}
then
\begin{eqnarray}
\CR(G, \theta_0, {\cal M})=
\CR(G_1, \theta_0, {\cal M}_1)+\CR(G_2, \theta_0, {\cal M}_2)
\nonumber
\end{eqnarray}
\label{lemma:independent}
\end{lemma}
When the premise of the lemma is satisfied,
${\cal M}_1$ and ${\cal M}_2$ are said to be 
{\it informationally independent} at $\theta_0$.
\begin{proof}
Let $M$ be a locally unbiased measurement  in ${\cal M}$,
and define 
the measurements $M_j\,(j=1,2)$ in $\R^{m_j}\,(j=1,2)$ by
\begin{eqnarray}  
M_1(B)&\equiv & M(B\times \R^{m_2})\:(B\in\R^{m_1}), \nonumber\\
M_2(B')&\equiv & M(\R^{m_1}\times B')\:(B'\in\R^{m_2}),\nonumber
\end{eqnarray}
respectively.
Then, the measurement $M_j\,(j=1,2)$ is  locally unbiased
in ${\cal M}_j\,(j=1,2)$, respectively.

Therefore, we have
\begin{eqnarray}
& &\inf\left\{\left.\Tr GV[M]\, \right|\, 
        \mbox{$M$ is locally unbiased in ${\cal M}$}\right\}\nonumber\\
&=&\inf\left\{\left.\Tr G_1 V[M_1]\,\right|\, 
        \mbox{$M$ is locally unbiased in ${\cal M}$}\right\}\nonumber\\
& &+\inf\left\{\left.\Tr G_2 V[M_2]\,\right|\, 
   \mbox{$M$ is locally unbiased in ${\cal M}$}\right\}\nonumber\\
&\ge& \inf\left\{\left.\Tr G_1 V[M']\,\right|\, 
        \mbox{$M'$ is locally unbiased in ${\cal M}_1$}\right\}\nonumber\\
& &+\inf\left\{\left.\Tr G_2 V[M'']\, \right|\, 
        \mbox{$M''$ is locally unbiased in ${\cal M}_2$}\right\}\nonumber
\end{eqnarray}
or its equivalence,
\begin{eqnarray}
\CR(G, {\cal M})\ge
\CR(G_1, {\cal M}_1)+\CR(G_2, {\cal M}_2).
\label{eqn:cr>cr+cr}
\end{eqnarray}
Because ${\cal M}_1$ and ${\cal M}_2$ are informationally independent,
${\sf L}$ for ${\cal M}$ writes
\begin{eqnarray}
{\sf L}=\left[
\begin{array}{cc}
{\sf L}_1 & 0 \\
0 & {\sf L}_2
\end{array}
\right],
\nonumber
\end{eqnarray}
in the appropriate coordinate,
where ${\sf L}_1=[|l_1\rgl,...,|l_{m_1}\rgl]$,
and ${\sf L}_2=[|l_{m_1+1}\rgl,...,|l_{m}\rgl]$.
In that coordinate,  let us write ${\sf X}$ as
\begin{eqnarray}
{\sf X}=\left[
\begin{array}{cc}
{\sf X}_1 & {\sf X}_{12}\\
{\sf X}_{21} & {\sf X}_2
\end{array}
\right].
\nonumber
\end{eqnarray}
Then
\begin{eqnarray}
{\rm Re} {\sf X}_i^*{\sf L}_i=I_{m_i},\,
{\rm Im}{\sf X}_i^*{\sf X}_i=0\,(i=1,2),\,
{\sf X}_{12}= {\sf X}_{21}=0,
\nonumber
\end{eqnarray}
is a sufficient condition for 
the measurements corresponding to ${\sf X}$ to be locally unbiased.
Therefore, we have
\begin{eqnarray}
& &\CR(G, {\cal M})\nonumber\\
&=&\inf\left\{\, \Tr G {\sf X}^*{\sf X}\,\left|\,
        {\rm Re} {\sf X}^*{\sf L}=I_m,\,{\rm Im}{\sf X}^*{\sf X}=0\right. 
        \,\right\}
\nonumber\\
&\leq&\inf\left\{\left. \sum_{i=1}^2\Tr G_i {\sf X}_i^*{\sf X_i}\, \right| \,
        {\rm Re} {\sf X}_i^*{\sf L_i}=I_{m_i},\,{\rm Im}{\sf X}_i^*{\sf X}_i =0,
                \,(i=1,2)\,\right\}
\nonumber\\
&=&\CR(G_1, {\cal M}_1)+\CR(G_2, {\cal M}_2),
\nonumber
\end{eqnarray}
which, mixed with $(\ref{eqn:cr>cr+cr})$ leads to the lemma.
\end{proof}

\section{Manifestation of complex structure}
\label{sec:complex}
It is worthy of notice that
$|\beta|$, which was shown to be 
a good index of `uncertainty' in the case of the $2$-dimensional model,
is deeply related to the natural complex structure in ${\cal P}_1$.

Let us define 
the linear transform  ${\bf D}$ in ${\cal T}_{\rho}({\cal M})$
as follows;
First, multiply the imaginary unit $i$ to $|l_X\rgl$.
In general, however,
$i|l_X\rgl$ is not an element of $span_{\bf R}{\sf L}$,
and  does not represent any of  vectors in ${\cal T}_{\rho}({\cal M})$.
Hence, we project $i|l_X\rgl$ onto  $span_{\bf R}{\sf L}$
with respect to the inner product ${\rm Re}\lgl *|* \rgl$,
and the image by $\pi_*$ of the product of the projection 
is defined to be ${\bf D}X\in {\cal T}_{\rho}({\cal M})$ ,
where $\pi_*$ is the differential map of $\pi$.

By elementary linear algebra, it is shown that
the matrix which corresponds to ${\bf D}$ is 
$J^{S-1}\tilde{J}$, and 
that, in the 2-dimensional model, its eigenvalues are $\pm i\beta$.

\begin{eqnarray}
\begin{array}{ccc}
        &\mbox{multiplication of}\:\: i&        
\\
|l\rgl  \in span_{\bf R}{\sf L}&--\longrightarrow
&i|l\rgl  \in span_{\bf R}\{{\sf L}, i{\sf L}\} 
\\
\uparrow& &\:\downarrow\mbox{project}
\\
\mbox{horizontal lift}& &|m\rgl\in span_{\bf R}{\sf L}
\\
\uparrow& &\downarrow\pi_*
\\
X\in{\cal T}_{\rho}({\cal M})&--\longrightarrow
& {\bf D}X\in{\cal T}_{\rho}({\cal M})
\\
  &{\bf D}&
\nonumber
\end{array}
\end{eqnarray}

The definition of the map ${\bf D}$ naturally
leads to the following theorems.
\begin{theorem}
The  absolute value
of the  eigenvalue of ${\bf D}$, or equivalently, of $J^{S-1}\tilde{J}$, 
is smaller than or equal to $1$.
\label{theorem:coherent:beta_i}
\end{theorem}

Is the eigenvalues
of the linear map ${\bf D}$
 a good measure of `uncertainty' in the arbitrary dimensional model? 
If all of eigenvalues of ${\bf D}$ vanish, 
as is shown in the section $\ref{sec:classical}$,
the model is quasi-classical, 
and `uncertainty' among parameters vanishes.
When  eigenvalues of ${\bf D}$ do not vanish,
  we have the following theorem.
\begin{theorem}
For any pure state model,
\begin{eqnarray}
&&\inf\{\Tr J^{S-1}V\:|\: V\in {\cal V}\}\nonumber\\
&=&\Tr
\left\{{\rm Re}\sqrt{I_m+i \sqrt{J^{S-1}}\tilde{J}
                            \sqrt{J^{S-1}}}\right\}^{-2}\nonumber\\
&=&\sum_{\beta\, :\,\mbox{eigenvalues of ${\bf D}$}} 
\frac{2}{1+\sqrt{1-|\beta|^2}}.
\nonumber
\end{eqnarray}
\end{theorem}
The estimation theoretical meaning 
of $\min\{\Tr J^{S-1}V\:|\: V\in {\cal V}\}$
is hard to verify. However, this value remains invariant under
any transform of the coordinate in the model ${\cal M}$,
and can be an index of distance between ${\cal V}$ and $J^{S-1}$.
\begin{proof} 
Because 
$\min\{\Tr J^{S-1}V\:|\: V\in {\cal V}\}$
is invariant by any affine coordinate transform in the model ${\cal M}$,
we choose a coordinate in which
$J^S$ writes $I_m$ and $\tilde{J}$ writes
\begin{eqnarray}
\tilde{J}=\left[
\begin{array}{ccccccccc}
0& -\beta_1&            0&\cdots&0&0&0&\cdots&0\\
\beta_1&0 &             0&\cdots&0&0&0&\cdots&0\\
0& 0&           \ddots&\ddots&\vdots&\vdots&\vdots&\ddots&\vdots\\
\vdots&\vdots&\ddots&\ddots&0&0&0&\cdots&0\\
0&0&            \cdots &0&0& -\beta_{l}&0&\cdots&0\\
0&0&            \cdots &0&\beta_{l}&0 &0&\cdots&0\\
0&0&            \cdots &0&    0    &0&0&\ddots&\vdots\\
\vdots&\vdots&\ddots&\vdots&\vdots&\vdots&\ddots&\ddots&\vdots\\
0&0&            \cdots &0&    0    &0 &\cdots&\cdots&0
\end{array}
\right]. 
\nonumber
\end{eqnarray}
Then, The model  ${\cal M}$ is decomposed into
the direct sum of the submodels one or two dimensional ${\cal M}_{\kappa}$,
\begin{eqnarray}
{\cal M}=\bigoplus_{\kappa} {\cal M}_{\kappa},
\nonumber
\end{eqnarray}
where 
any two submodels  ${\cal M}_{\kappa}$ and ${\cal M}_{\kappa'}$
are informationally independent, and 
$\tilde{J}$ of a two dimensional submodel ${\cal M}_{\kappa}$
is 
\begin{eqnarray}
\left[
\begin{array}{cc}
0& -\beta_{\kappa} \\
\beta_{\kappa}&0 
\end{array}
\right].
\nonumber
\end{eqnarray}
Therefore, by virtue of lemma $\ref{lemma:independent}$ 
and the equation $(\ref{eqn:minvv})$,
we have the theorem.
\end{proof}

\section{The coherent model}
\label{sec:coherent}
As for the model with arbitrary dimensions,
the model is said to be {\it coherent} at $\theta$
iff all of the eigenvalues of $(J^S)^{-1}\tilde{J}$ are $\pm i$.
When the model is $2$-dimensional,
this definition of coherency reduces to $|\beta|=1$.
The dimension of the coherent model is  even,
for the eigenvalues of $J^{S -1}\tilde{J}$ are 
of the form $\pm i\beta_j$ or 0.

In this section, we determine the attainable CR type bound of 
the coherent model.
The coherent model is worthy of attention firstly 
because the coherent model is `the maximal uncertainty' model,
secondly because there are several physically important models 
which are coherent.

The definition of the map ${\bf D}$ leads to
the following theorem.
\begin{lemma}
The model ${\cal M}$ is coherent at $\theta$ iff 
$span_{\R}\{i{\sf L}\}$ is identical to $span_{\R}{\sf L}$,
or equivalently, iff 
$span_{\R}\{ {\sf L}, i{\sf L}\}$ is identical to 
$span_{\R}{\sf L}$.
\label{lemma:spaniL}
\end{lemma}
This lemma  leads to the following lemma.
\begin{lemma}
The model $\cal M$ is coherent  iff
the dimension of $span_{\C}{\sf L}$ is $m/2$.
\end{lemma}
\begin{proof}
First, we  assume that 
\begin{eqnarray}
\dim_{\C}span_{\C}{\sf L}=m/2.
\label{eqn:dimL=m/2}
\end{eqnarray}
Because $span_{\R}{\sf L}$ is a $m$-dimensional subspace of 
$span_{\R}\{{\sf L},i{\sf L}\}$ whose dimension is
smaller than or equal to $m$ because of $(\ref{eqn:dimL=m/2})$,
we have 
$span_{\R}\{{\sf L},i{\sf L}\}=span_{\R}{\sf L}$, 
or coherency of the model.

Conversely, let us assume that the model is coherent.
Taking an orthonormal basis $\{e_j|\: j=1,...m\}$ of 
$span_{\R}{\sf L}$ such that
$e_{j+m/2}={\bf D}e_j$, horizontal lifts $\{|j\rgl |\: j=1,...m\}$ of
$\{e_j|\: j=1,...m\}$ satisfy $|j+m/2\rgl=i|j\rgl$,
and any element $|u\rgl$ of  
$span_{\R}{\sf L}= span_{\R}\{{\sf L}, i{\sf L}\}$ 
writes
\begin{eqnarray}
|u\rgl &=&\sum_{j=1}^m a_j |j\rgl \nonumber\\
&=&\sum_{j=1}^{m/2} (a_j+ i a_{j+(m/2)})|j\rgl,\nonumber
\end{eqnarray}
implying that the dimension of $span_{\C}{\sf L}$ is $m/2$.
\end{proof}

Fujiwara and Nagaoka \cite{FujiwaraNagaoka:1996}
determined the attainable CR type bound of 
the two parameter coherent model.
In the following, more generally, 
we work on the bound of the coherent model with arbitrary
dimension. 
Throughout the section, the weight matrix $G$ 
is assumed to be strictly positive.

When the model is coherent, ${\rm Re}{\sf L}^*{\sf X}=I_m$ 
or equivalently ${\rm Re}{\sf L}^*({\sf X}-{\sf L}J^{S-1})=0$,
implies, by virtue of $span_{\R}{\sf L}=span_{\R}\{i{\sf L},\,{\sf L}\}$,
\begin{eqnarray}
{\sf L}^*({\sf X}-{\sf L}J^{S-1})=0,
\nonumber
\end{eqnarray}
or equivalently,
\begin{eqnarray}
{\sf L}^*{\sf X}={\sf L}^*{\sf L}J^{S-1}=I_m+i{\tilde J}J^{S-1}.
\label{eqn:lx=i+ijj}
\end{eqnarray}

Multiplication of  ${\sf L}^*$ to the both sides of $(\ref{eqn:basic0.1})$,
together with the equation $(\ref{eqn:lx=i+ijj})$, yields
\begin{eqnarray}
(I_m+i{\tilde J}J^{S-1})(G-i\Lambda)=(J^S+i\tilde{J})VG.
\label{eqn:coherent:g-il}
\end{eqnarray}
By virtue of the coherency,
both of the real part and the imaginary part of 
$(\ref{eqn:coherent:g-il})$ give the same equation,
\begin{eqnarray}
G+{\tilde J}J^{S-1}\Lambda=J^SVG,
\nonumber
\end{eqnarray}
or
\begin{eqnarray}
\sqrt{G}V\sqrt{G}-
\sqrt{G}J^{S-1}\sqrt{G}=
\left(\sqrt{G}J^{S-1}{\tilde J}J^{S-1}\sqrt{G}\right)
\left(G^{-1/2}\Lambda G^{-1/2}\right).
\label{eqn:gvg-gjg}
\end{eqnarray}
The antisymmetric part of the both hands of the equation
yields
\begin{eqnarray}
\left[\sqrt{G}J^{S-1}{\tilde J}J^{S-1}\sqrt{G},
       \: G^{-1/2}\Lambda G^{-1/2}\right]=0.
\nonumber
\end{eqnarray}

Therefore, letting $a_i$ and $b_i$ denote  the eigenvalues of 
$\sqrt{G} J^{S-1}{\tilde J}J^{S-1}\sqrt{G}$ and
$G^{-1/2}\Lambda G^{-1/2}$ respectively, we have
\begin{eqnarray}
&&\Tr \left(\sqrt{G}J^{S-1}{\tilde J}J^{S-1}\sqrt{G}\right)
              \left(G^{-1/2}\Lambda G^{-1/2}\right)\nonumber\\
&=& \sum_i a_i b_i =\sum_i |a_i||b_i|,
\nonumber
\end{eqnarray}
where the last equality is valid because
the left hand side of the equation $(\ref{eqn:gvg-gjg})$
is positive symmetricity virtue of the SLD CR inequality.

On the other hand, from $(\ref{eqn:basic0.1})$ or its equivalence,
\begin{eqnarray}
{\sf X}\sqrt{G}\left(I_m-iG^{-1/2}\Lambda G^{-1/2}\right)
={\sf L}V\sqrt{G},
\label{eqn:coherent:i-iglg}
\end{eqnarray}
we can deduce 
$|b_i|=1\,(i=1,...,m)$ as in the follows.

The rank of the right hand side of $(\ref{eqn:coherent:i-iglg})$
is equal to $m/2$ because $G$ is strictly positive and
\begin{eqnarray}
\rank {\sf L}=\dim_{\C} span_{\C} {\sf L}=m/2.
\nonumber
\end{eqnarray}
On the other hand, 
\begin{eqnarray}
\rank{\sf X}=\dim_{\C} span_{\C}{\sf X}= \dim_{\R} span_{\R}{\sf X} =m,
\nonumber
\end{eqnarray}
where the second equality comes from ${\rm Im}{\sf X}^*{\sf X}=0$
and the last equality comes from ${\rm Re}{\sf X}^*{\sf L}=I_m$.
Therefore, the rank of the matrix $I_m-iG^{-1/2}\Lambda G^{-1/2}$ 
must be $m/2$, and the eigenvalues of $G^{-1/2}\Lambda G^{-1/2}$
are $\pm i$.

After all, we have
\begin{eqnarray}
\min_{V\in{\cal V}({\cal M})}\Tr GV
=\Tr GJ^{S-1}+\Tr\abs GJ^{S-1}{\tilde J}J^{S-1},
\nonumber
\end{eqnarray}
where $\Tr\abs A$ means 
the sum of the absolute values of the eigenvalues of the matrix $A$.
When the minimum is attained, the covariance matrix $V$ is given by
\begin{eqnarray}
V=J^{S-1}+G^{-1/2}|G^{1/2}J^{S-1}{\tilde J}J^{S-1}G^{1/2}|G^{-1/2},
\nonumber
\end{eqnarray}
where $|A|=(AA^*)^{1/2}$.

To check the coherency of the model, the following theorem,
which is deduced from 
theorem $\ref{theorem:coherent:beta_i}$, is
useful.
\begin{theorem}
the model is coherent at $\theta$
iff 
\begin{eqnarray}
|{\rm det} J^S|=|{\rm det} \tilde{J}|.
\nonumber
\end{eqnarray}
\label{theorem:coherent:check}
\end{theorem}
\begin{example}(squeezed state model)
{\it Squeezed state model}, which has four parameters, is defined by
\begin{eqnarray}
{\cal M}=\{\rho(z, \xi)\:|\; \rho(z,\xi)=|z,\xi\rgl\lgl z,\xi|,\; z,\xi\in\C\},
\nonumber
\end{eqnarray}
where
\begin{eqnarray}
|z,\xi\rgl&=&D(z)S(\xi)|0\rgl,\nonumber\\
D(z)&=&\exp(za^{\dagger}-\ol{z} a),\nonumber\\
S(\xi)&=&\exp\left(\frac{1}{2}(\xi a^{\dagger2}-\ol{\xi} a^2)\right).
\nonumber
\end{eqnarray}
Letting $z=(\theta^1+i\theta^2)/2^{1/2}, Q=(a+a^{\dagger})/2^{1/2},$
and $\xi=\theta^3 e^{-2i\theta^4}\;(0\leq \theta^3,\, 0\leq\theta^4 <2\pi)$,
we have
\begin{eqnarray}
J^S&=&\frac{1}{2}\left[
\begin{array}{cccc}
\cosh 2\theta^3-\sinh 2\theta^3 \cos 2\theta^4 &
\sinh 2\theta^3\sin 2\theta^4 &0&0\\
\sinh 2\theta^3\sin 2\theta^4 & 
\cosh 2\theta^3+\sinh 2\theta^3 \cos 2\theta^4 &0&0\\
0&0&1&0\\
0&0&0&\sinh^2 2\theta^3
\end{array}
\right],\nonumber\\
\tilde{J}&=&\frac{1}{2}\left[
\begin{array}{cccc}
0&1&0&0\\
-1&0&0&0\\
0&0&0&-\sinh 2\theta^3\\
0&0&\sinh 2\theta^3&0
\end{array}
\right]
\nonumber
\end{eqnarray}

Coherency of this model is easily checked 
by theorem \ref{theorem:coherent:check},
\begin{eqnarray}
|{\rm det} J^S|=|{\rm det} \tilde{J}|=\frac{1}{4}\sinh^2 2\theta^3.
\nonumber
\end{eqnarray}
\end{example}
\begin{example}(spin coherent model)
As is pointed out by Fujiwara $\cite{FujiwaraNagaoka:1996}$,
{\it spin coherent model}${\cal M}_{s,s}$,
which is a special case of spin rotation model $(\ref{eqn:gscoherent})$,
is coherent.
\end{example}
\begin{example}(total space model)
 {\it The total space model} is the space of all the pure state ${\cal P}$
in finite dimensional Hilbert space ${\cal H}$.
By virtue of theorem \ref{lemma:spaniL},
the coherency of the model is proved by
checking that $span_{\R}{\sf L}$ is invariant by the multiplication of 
the imaginary unit $i$.
Let $|l\rgl$ be a horizontal lift of a tangent vector at $|\phi\rgl$.
Then, $i|l\rgl$ is 
also a horizontal lift of another tangent vector at $|\phi\rgl$,
because $|\phi\rgl+i|l\rgl dt$ is an element of ${\cal H}$ with unit length.
\end{example}

\section*{Acknowledgement}
The author is
grateful to Dr. A. Fujiwara and Dr. K. Nagaoka for inspiring discussions.
The author is indebted to Mr. M. Hayashi for pointing out
the fact that  the attainable CR bound is attained 
asymptotically in the sense stated 
in the end of the section $\ref{sec:unbiased}$.

\appendix
\section{proof of lemma 3}
\begin{proof}
Let $E^{(i)}$ be a projection valued measure such that,
\begin{eqnarray}
\int_{\R} x E^{(i)}(dx)
=\sum_{j=1}^{m}\left [J^{S-1}(\theta)\right]^{ij}L^S_j(\theta),
\nonumber
\end{eqnarray}
and $M_{\bf p}$ be an unbiased measurement at $\theta$ such that
\begin{eqnarray}
&&M_{\bf p}\left(\left\{\theta^1\right\}\times ...
    \times 
    \left[ \theta^i+\frac{x}{p_i},\,\theta^i+\frac{x+\Delta x}{p_i}\right] 
   \times ... \times\left\{\theta^{m}\right\}\right)
=p_i E^{(i)}([x,\, x+\Delta x]),\nonumber\\
&&M_{\bf p}\left( B_1\times ...
                  \times \left(\R/\left\{ \theta^i \right\}\right) 
                  \times
                    ... \times B_m\right)=0
\nonumber
\end{eqnarray}
where ${\bf p}=[p_i]\,(i=1,...m)$ is a real vector such that
$\sum_i p_i=1$ and $p_i\geq 0$, and $B_i\,(i=1,...,m)$ are 
arbitrary measurable subset of $\R$.
Then, we have for any ${\bf p}$,
\begin{eqnarray}
[V_{\theta}[M_{\bf p}]]_{ii}= \frac{1}{p_i}\left[J^{S-1}(\theta)\right]^{ii},
\nonumber
\end{eqnarray}
which leads to 
\begin{eqnarray}
&&\inf \left\{\left. [V_{\theta}[M]]_{ii}\, \right|\, 
          \mbox{ $M$ is locally unbiased at $\theta$}\right\}
\nonumber\\
&\leq& \inf\left\{[V_{\theta}[M_{\bf p}]]_{ii}\, \left|\, \sum_i p_i=1,\,p_i\geq 0
                                                      \right.\right\}
\nonumber\\
&=&\left[J^{S-1}(\theta)\right]^{ii}.
\nonumber
\end{eqnarray}
On the other hand, SLD CR inequality leads to
\begin{eqnarray}
\inf \left\{\left. [V_{\theta}[M]]_{ii}\, \right|\, 
          \mbox{ $M$ is locally unbiased at $\theta$}\right\}
\geq \left[J^{S-1}(\theta)\right]^{ii},
\nonumber
\end{eqnarray}
and we have the lemma.
\end{proof}

\section{proof of lemma 5}
\begin{proof}
Let $M$ be a locally unbiased measurement, and
${\bf v}_{\bf \alpha}=(v^{\bf \alpha}_1,...,v^{\bf \alpha}_m) $ 
denote $\sqrt{V_0}\alpha$,
where $\alpha$ is a vector whose components are $1$ or $-1$.
Then, the measurement $M'$, which is  defined by
\begin{eqnarray}
M'\left(\prod_{i=1}^m
  ([a_i - v_{\bf\alpha}^i,\,\, b_i  - v_{\bf \alpha}^i]
  \cup [a_i + v_{\bf\alpha}^i ,\,\,  b_i + v_{\bf\alpha}^i\,\right)
=\frac{1}{2^m} M \left(\prod_{i=1}^m [a_i,\,\,  b_i ]\right)
\nonumber
\end{eqnarray}
is also locally unbiased and its covariance matrix is,
\begin{eqnarray}
V[M']=V[M]+V_0.
\nonumber
\end{eqnarray}
\end{proof}

\section{proof of lemma 7}
\begin{proof}
The equation $(\ref{eqn:imxx=0})$ and 
the equation  $(\ref{eqn:naimark:unbiased})$
implies that, for any element $V$ of ${\cal V}$,
there is a $m\times 2m$ matrix $U$ which satisfies
\begin{eqnarray}
V^{-1}=V^{-1}(U)\equiv({\rm Re} U{\sf L})^T{\rm Re} U{\sf L},
\nonumber
\end{eqnarray}
and
\begin{eqnarray}
U^*U=I_m.
\label{eqn:uu=1}
\end{eqnarray}
Because the map $V^{-1}(*)$ is continuous
and the totality of the $m\times 2m$ matrix $U$ 
satisfying $(\ref{eqn:uu=1})$ 
is compact, the region of $V^{-1}(U)$ is compact.
Therefore,
the intersection of ${\cal V}$ and the set
\begin{eqnarray}
\{ A \: |\:  A\leq V_0\}.
\nonumber
\end{eqnarray}
is compact for any real symmetric matrix $V_0$,
for 
the map
\begin{eqnarray}
V^{-1}\rightarrow V
\nonumber
\end{eqnarray}
is continuous on the intersection of  the region of $V^{-1}(U)$
and the set
\begin{eqnarray}
\{ A \: |\:  A\geq V^{-1}_0\},
\nonumber
\end{eqnarray}
both of which are compact.
Because $V_0$ is an arbitrary real symmetric matrix,
we have the lemma.
\end{proof}

\end{document}